\documentclass[12pt]{article}
\usepackage{epsfig}
\usepackage[hyperindex=true]{hyperref}
\usepackage{epsfig}

\newcommand{\ii}{\mathrm{i}}
\newcommand{\ee}{\mathrm{e}}
\newcommand{\dd}{\mathrm{d}}

\newcommand{\sn}{\mathrm{sn}}
\newcommand{\dn}{\mathrm{dn}}
\newcommand{\cn}{\mathrm{cn}}
\newcommand{\Z}{\mathrm{Z}}
\newcommand{\K}{\mathbf{K}}
\newcommand{\E}{\mathbf{E}}
\newcommand{\EK}{\frac{\E}{\K}}
\newcommand{\re}{\mathrm{Re}}
\newcommand{\im}{\mathrm{Im}}
\newcommand{\w}{\omega}
\newcommand{\kk}{\kappa}
\newcommand{\e}{\varepsilon}
\begin{document}
\thispagestyle{empty}
\noindent\hspace*{\fill} FAU-TP3-05/6\\

\begin{center}\begin{Large}\begin{bf}
Full Phase Diagram of the Massive\\ Gross-Neveu Model
\\
\end{bf}\end{Large}\vspace{.75cm}
\vspace{0.5cm} Oliver Schnetz, Michael Thies and Konrad Urlichs\\ Institut f\"ur Theoretische Physik III \\
Universit\"at Erlangen-N\"urnberg, Erlangen, Germany\\
\vspace{1cm}\baselineskip=35pt
\end{center}
\date{November 28, 2005}
\begin{abstract}
The massive Gross-Neveu model is solved in the large $N$ limit at finite temperature and chemical potential.
The scalar potential is given in terms of Jacobi elliptic functions.
It contains three parameters which are determined by transcendental equations. Self-consistency of the
scalar potential is proved. The phase diagram for non-zero bare quark mass is found to contain a kink-antikink crystal
phase as well as a massive fermion gas phase featuring a cross-over from light to heavy effective fermion mass.
For zero bare quark mass we recover the three known phases kink-antikink crystal, massless fermion gas,
and massive fermion gas. All phase transitions are shown to be of second order. Equations for the phase boundaries are
given and solved numerically. Implications on condensed matter physics are indicated where our results
generalize the bipolaron lattice in non-degenerate conducting polymers to finite temperature.
\end{abstract}

\newpage

\section{Introduction and motivation}
In its original form, the Gross-Neveu (GN) model \cite{1} is a relativistic, renormalizable quantum field theory of $N$
species of self-interacting fermions in 1+1 dimensions with Lagrangian
\begin{equation}
\label{eq:1}
{\cal L} = \sum_{i=1}^N \bar{\psi}^{(i)} ({\rm i}\gamma^{\mu}\partial_{\mu} - m_0)\psi^{(i)} + \frac{1}{2} g^2 \left(\sum_{i=1}^N
\bar{\psi}^{(i)}\psi^{(i)}\right)^2.
\end{equation}
The common bare mass term $\propto m_0$ explicitly breaks the discrete chiral symmetry $\psi\to \gamma^5 \psi$ of the
massless model.
As far as the phase diagram is concerned, the 't Hooft limit $N\to \infty$, $g^2 \propto 1/N$ is most interesting
since it allows one to bypass some of the limitations of low dimensions, while justifying a semi-classical approach.
Interest from the particle physics side stems from the fact that the simple Lagrangian (\ref{eq:1}) shares
non-trivial properties with quantum chromodynamics (QCD), notably asymptotic freedom,
dimensional transmutation, meson and baryon bound states, chiral symmetry breaking in the vacuum as well as its
restoration
at high temperature and density (for a pedagogical review, see \cite{2}). It also has played some role as a testing ground for
fermion
algorithms on the lattice \cite{3}.
Perhaps even more surprising and less widely appreciated is the fact that GN type models have enjoyed
considerable success in describing a variety of
quasi-one-dimensional condensed matter systems, ranging from the Peierls-Fr\"ohlich model \cite{4} over ferromagnetic
superconductors \cite{5} to conducting polymers, e.g. doped {\em trans}-polyacetylene \cite{6}. By way of example, the kink
and kink-antikink
baryons first derived in field theory in the chiral limit ($m_0=0$) \cite{7} have been important for understanding the role of
 solitons
and polarons in
electrical conductivity properties of doped polymers. Likewise, one can show that baryons in the massive
 GN model
($m_0 \neq 0$) are closely related to polarons and bipolarons in polymers with non-degenerate ground states, e.g.
{\em cis}-polyacetylene
\cite{8,9,9a}. Finally, GN models with finite $N$ have been found useful for describing electrons in carbon nanotubes
and fullerenes \cite{10,10a}.

The claim that a relativistic theory is relevant for condensed matter physics is at first sight very provocative \cite{11}. A closer
inspection shows that a continuum approximation to a discrete system, a (nearly) half-filled band and a linearized dispersion
relation of the electrons at the Fermi surface are the crucial ingredients leading to a
Dirac-type theory. The Fermi velocity plays the role of the velocity of light and the band width the
 role of the ultra-violet (UV) cutoff.
In some cases the correspondence is literal so that results can be taken over from one field into the other.
This was illustrated in \cite{12} where we borrowed results from the theory of non-degenerate conducting
polymers, in particular the bipolaron lattice, for solving the zero temperature limit of the massive GN model.

In the process of renormalization the bare mass $m_0$ of the GN model is replaced by a physical constant
$\gamma \geq 0$, whereas
the relation between the coupling constant and the cutoff merely sets the scale of the theory. As a result the massive
 GN model
can be considered as a one parameter family of physical theories, the point $\gamma=0$ corresponding to the massless
 (chiral) case.
In this paper we determine the full phase diagram of the massive GN model as a function of
temperature $T$ and chemical potential $\mu$ for any value of $\gamma$. This issue was addressed by
Barducci et al.\ in 1995 \cite{13} (for earlier, partial results, see also \cite{13a}). Recent findings about the massless GN
 model \cite{14,15,16} cast doubts on the full validity of their
calculations. The assumption that the scalar condensate is spatially homogeneous is apparently too restrictive
and misses important physics related to the existence of kink-antikink baryons.
As we now know, the phase diagram in Ref.~\cite{13} has neither the correct
limit $m_0\to 0$ \cite{16}, nor the correct limit $T\to 0$ \cite{12}.
In view of the simplicity and generic character of the GN model, its potential for applications to different fields of physics,
and the fact that it was conceived more than three decades ago, it seems desirable to resolve these contradictions.

The key to the solution of the problem is twofold.
First we conjecture the expression for the self-consistent scalar potential.
To this end a three parameter ansatz will be presented and discussed in Sect.~2.
Second we formulate the grand potential in a way suitable for analytical calculations.
In Sect.~3 we show that the grand potential fits into the structure of complex analysis.
Minimizing it leads to three transcendental relations connecting the parameters of our ansatz to the
physical quantities $T$, $\mu$, and $\gamma$. From these equations it will be possible to read off self-consistency.
We outline an efficient procedure to numerically solve the parameter equations and present the full phase diagram in Sect.\ 4.
A detailed comparison with previous results will be given. We prove that all phase transitions are second order and
investigate singular points on the phase boundary. A Ginzburg-Landau effective theory valid in the vicinity of
the multicritical point is deduced and finally the zero temperature limit will be discussed.
Sect.~5 contains a brief summary and outlook. In Appendix A we collect some technicalities
needed for the proof of self-consistency, whereas Appendix B contains the basic formulae necessary for numerical
calculations. A short
version of this paper without detailed formalism has been presented previously \cite{17}.

\section{Generalized scalar potential for the massive Gross-Neveu model}
A convenient framework for studying the thermodynamics of the GN model is provided by the
relativistic Hartree-Fock approximation, expected to become rigorous in the large-$N$ limit.
Due to the $(\bar{\psi}\psi)^2$-interaction in the Lagrangian (\ref{eq:1}), the Dirac-Hartree-Fock equation assumes
the form
\begin{equation}
\label{eq:2}
\left( -{\rm i} \gamma^5   \frac{\partial}{\partial x} + \gamma^0 S(x) \right)\psi(x)=E \psi(x)
\end{equation}
with real scalar potential $S(x)$.
The scalar potential is related to the thermal expectation value of $\bar{\psi}\psi$ [see Eq.~(\ref{eq:54})].
It thus depends on the eigenfunctions $\psi$ and eigenvalues $E$ of (\ref{eq:2}) through a self-consistency relation.
At present $S(x)$ is regarded as an external potential acting on the fermions; self-consistency will be proved in Sect.~3.
We choose the following representation of the $\gamma$-matrices,
\begin{equation}
\gamma^0 = - \sigma_1 , \quad \gamma^1={\rm i}\sigma_3, \quad \gamma^5=\gamma^0\gamma^1 = - \sigma_2 .
\label{eq:3}
\end{equation}
In this representation,
the equations for the upper and lower components $\phi_\pm$ of the Dirac spinor $\psi$ can
be decoupled by squaring Eq.\ (\ref{eq:2}),
\begin{equation}
\label{eq:4}
\left( -  \frac{\partial^2}{\partial x^2} \mp  \frac{\partial}{\partial x} S+ S^2\right) \phi_{\pm} = E^2 \phi_{\pm} .
\end{equation}
Eq.~(\ref{eq:4}) states that the Schr\"odinger-type Hamiltonians with potentials $U_{\pm}=S^2 \pm S'$ have the same spectra,
a textbook example of supersymmetric quantum mechanics with superpotential $S$.
Guided by all the special cases already known,
we shall use the following ansatz for $S(x)$ depending on three real parameters
$A,\kk,b$
\begin{eqnarray}
&&S(x) = A\tilde{S}(\xi),\quad \xi =A x  ,\nonumber\\
&&\tilde{S}(\xi) = \kk^2 \sn\, b\, \sn\, \xi \, \sn(\xi+b) + \frac{\cn\, b\, \dn\, b}{\sn\, b}.
\label{eq:5}
\end{eqnarray}
Here $\sn$, $\cn$, and $\dn$ are Jacobi elliptic functions of modulus $\kk$ which will be suppressed throughout the paper.
Since this ansatz is central for the following, we give a detailed discussion of its properties as well as its
physical origin.

\vskip 0.1cm
{\em 1) Symmetries}
\vskip 0.1cm
As a real function, $\tilde{S}$ has a discrete translational symmetry in $\xi$ and $b$ with period $2{\K}$
(in this paper ${\K},{\bf E}$ denote complete elliptic integrals of first and second kind,
${\K'},{\bf E'}$ the complementary ones with argument $\kk'=\sqrt{1-\kk^2}$),
\begin{equation}
\tilde{S}(\xi,b)=\tilde{S}(\xi + 2 {\K},b)=\tilde{S}(\xi,b + 2 {\K}) .
\label{eq:6}
\end{equation}
Moreover, it is symmetric under reflections about $\xi=-b/2$ and about $\xi=\K-b/2$ as well as antisymmetric under
simultaneous reflections
of $\xi$ and $b$,
\begin{equation}
\tilde{S}(\xi,b)=\tilde{S}(-b-\xi,b)=-\tilde{S}(-\xi,-b)  .
\label{eq:7}
\end{equation}
From the Dirac-Hartree-Fock equation (\ref{eq:2}) we see that $S(x)$ and $-S(-x)$ are equivalent.
Therefore we find that the space of different configurations is parameterized by
\begin{equation}
A\geq 0,\quad 0\leq b\leq {\K},\quad 0\leq \kk \leq 1 .
\label{eq:8}
\end{equation}
In the complex $\xi$-plane $\tilde{S}$ has a second period of $2\ii {\K'}$,
\begin{equation}
\tilde{S}(\xi,b)=\tilde{S}(\xi + 2\ii {\K'},b) .
\label{eq:9}
\end{equation}
This, together with the fact that it is meromorphic, makes $\tilde{S}$ an elliptic function in $\xi$. (It is elliptic in $b$, too, but
we will not need this here.) As an elliptic function it becomes worthwhile to study $\tilde{S}$ in the complex plane.

\vskip 0.1cm
{\em 2) Special values}
\vskip 0.1cm
From
\begin{equation}
\tilde{S'}(\xi)=\kk^2\sn\, b \, \sn\,(2\xi+b)   \left[1-\kk^2\sn^2\xi\,\sn^2(\xi+b)\right]
\label{eq:10}
\end{equation}
we see that on the real line $\tilde{S}$ has its extrema at the symmetric points $-b/2+n\K$; minima for even $n$ and maxima
for odd $n$.
In the complex $\xi$-plane $\tilde{S}$ is of degree two: It has two poles (or one double pole) in its fundamental domain.
We find
\begin{eqnarray}
&&\tilde{S}(-b/2)\;=\;\tilde{S}_{\rm min}\;=\;\frac{\cn\,b+\dn\,b-1}{\sn\,b}\nonumber\\
&&\tilde{S}(\K/2-b/2)\;=\;\tilde{S}(3\K/2-b/2)\;=\;\frac{\dn\,b-\kk'\,\sn^2b}{\sn\,b\,\cn\,b}\nonumber\\
&&\tilde{S}(\K-b/2)\;=\;\tilde{S}_{\rm max}\;=\;\frac{\cn\,b-\dn\,b+1}{\sn\,b}\nonumber\\
\label{eq:11}
&&\tilde{S}(\ii\K')\;=\;\tilde{S}(-b+\ii\K')\;=\;\infty  .
\end{eqnarray}
The Laurent series at $\xi=\ii\K'$ starts with the coefficients
\begin{equation}
c_{-1}=\hbox{res}_{\xi=\ii\K'}\tilde{S}=1,\quad c_0=0,\quad c_1=\frac{1}{\sn^2\, b}-\frac{1+\kk^2}{3} .
\label{eq:12}
\end{equation}
The Laurent coefficients $c'_k$ at $\xi=-b+\ii\K'$ are determined by reflection symmetry $c'_k=(-1)^k c_k$.

\vskip 0.1cm
{\em 3) Alternative expressions}
\vskip 0.1cm
The expression (\ref{eq:5}) for $\tilde{S}$ can be cast into various alternative forms. Any of these expressions can be easily
checked by comparing its pole structure with Eq.~(\ref{eq:12}). The boundedness theorem of complex analysis guarantees that
two elliptic
functions of equal periods are identical up to an additive constant once their pole structure is the same.

The symmetrized $\tilde{S}(\xi-b/2)$ can be expressed in terms of the Weierstrass
elliptic function ${\cal P}(z)$ with periods $\w_1=2\ii\K'$, $\w_2=2\K$ (see \cite{18a}, $\xi_{\pm}=\xi\pm b/2$),
\begin{eqnarray}
\tilde{S}(\xi_-)&=&\frac{1}{2}
\frac{{\cal P}'(\xi_+ +\ii\K')+{\cal P}'(\xi_- +\ii\K')}{{\cal P}(\xi_+ +\ii\K')-{\cal P}(\xi_- +\ii\K')}\nonumber\\
&=&\frac{{\cal P}'(b/2+\ii\K')}{{\cal P}(\xi)-{\cal P}(b/2+\ii\K')}+\frac{{\cal P}''(b/2+\ii\K')}{2{\cal P}'(b/2+\ii\K')},
\label{eq:13}
\end{eqnarray}
or as ratio of Jacobi elliptic functions \cite{21},
\begin{equation}
\label{eq:14}
\tilde{S}(\xi_-) = \frac{\sn\,\xi_{+} \cn\, \xi_{+} \dn\, \xi_{+} + \sn\,\xi_{-} \,\cn\,\xi_{-} \,\dn\,\xi_{-}}{\sn^2\xi_{+}-\sn^2\xi_{-}} .
\end{equation}
Other alternatives for $\tilde{S}(\xi)$ are
\begin{eqnarray}
\tilde{S}(\xi) &=& -\frac{\kk^2 \sn\, b}{\dn\, b}\, \cn\, \xi \, \cn(\xi+b) + \frac{\cn\, b}{\sn\, b\, \dn\, b}\nonumber\\
&=& -\frac{\sn\, b}{\cn\, b}\, \dn\, \xi \, \dn(\xi+b) + \frac{\dn\, b}{\sn\, b\, \cn\, b} .
\label{eq:15}
\end{eqnarray}
Moreover, note that
\begin{equation}
\tilde{S}(\xi)^2 = \frac{1}{\sn^2 b}+1-\kk^2-\dn^2 \xi-\dn^2(\xi+b).
\label{eq:16}
\end{equation}

\vskip 0.1cm
{\em 4) Limits}
\vskip 0.1cm
The elliptic modulus $\kk$ varies between 0 and 1. At the boundaries of this interval, we find
\begin{eqnarray}
\lim_{\kk \to 0} \tilde{S}(\xi) &=& \cot b,
\nonumber \\
\lim_{\kk \to 1 } \tilde{S}(\xi) & = & \coth b + \tanh \xi - \tanh (\xi+b) .
\label{eq:17}
\end{eqnarray}
The first limit is a constant, the second represents a single baryon profile. The  limits  $\kk \to 0$ and $\kk \to 1$ will
later describe the transition from the crystal phase with periodic $S$ to the massive Fermi gas phase with constant $S$
and mass $A\cot b$ or $A\coth b$, respectively. Our ansatz for $S$ is capable of parameterizing
both a periodic crystal and a spatially homogeneous phase.

The limiting values for $b$ are $b=0$ and $b=\K$ where we find
\begin{eqnarray}
\tilde{S}(\xi) &=& 1/b+{\cal O}(b)\quad\hbox{and}
\nonumber \\
\lim_{b \to {\K}} \tilde{S}(\xi) &=& \kk^2\, \frac{\sn\, \xi \,\cn\, \xi}{\dn\, \xi}.
\label{eq:18}
\end{eqnarray}
Zero is the only value of $b$ where $\kk=0$ and $\kk=1$ can coexist and will later be a singular point
in the phase diagram. The limit $b\to \K$ yields the form of the self-consistent
potential familiar from the chiral limit \cite{16}.

\vskip 0.1cm
{\em 5) Hamiltonian potentials}
\vskip 0.1cm
From the singularity structure of $\tilde{S}$, Eq.~(\ref{eq:12}), we read off that the Hamiltonian potentials
$U_{\pm}(x)=A^2\tilde{U}_{\pm}(\xi)$, $\tilde{U}_{\pm}=\tilde{S}^2 \pm \tilde{S}'$ have only one pole which is of second order.
For $\tilde{U}_-$ the singularity at $\xi=\ii\K'$ remains and we obtain the Laurent coefficients
\begin{equation}
c_{-2}=2,\quad c_{-1}=\hbox{res}_{\xi=\ii\K'}\tilde{U}_-=0,
\quad c_0=\frac{1}{\sn^2\, b}-\frac{1+\kk^2}{3}.
\label{eq:19}
\end{equation}
Moreover $\tilde{U}_-(\xi)$ is an elliptic function and shares periods and poles with $2\kk^2\sn^2\xi$. Comparison
 with $c_0$ in the
above expansion immediately yields
\begin{equation}
\tilde{U}_-(\xi)=\tilde{S}(\xi)^2-\tilde{S}(\xi)'=2\kk^2\sn^2 \xi+\frac{1}{\sn^2\, b}-1-\kk^2 .
\label{eq:20}
\end{equation}
The result for $\tilde{U}_+$ is obtained by reflection symmetry $\tilde{U}_+(\xi)=\tilde{U}_-(-b-\xi)$.

\vskip 0.1cm
{\em 6) Spatial averages}
\vskip 0.1cm
We need the spatial averages of $\tilde{S}$ and $\tilde{S}^2$ in the following sections. While the calculation of
the latter is straightforward using either Eq.~(\ref{eq:16}) or Eqs.~(\ref{eq:41}, \ref{eq:45}),
the first one is more intricate. To derive it we have used the Fourier
decomposition of the Jacobi elliptic functions to get
\begin{equation}
\frac{1}{2\K}\int_0^{2\K}{\rm d }\xi \,\sn\,\xi\,\sn\,(\xi+b)=\frac{2\pi^2}{\kk^2\K^2}\sum_{n=0}^\infty
\frac{q^{2n+1}}{(1-q^{2n+1})^2}\cos
\frac{(2n+1)\pi b}{2\K} ,
\label{eq:21}
\end{equation}
where $q=\exp(-\pi\K'/\K)$ is the elliptic nome of $\kk$. Upon comparing the Fourier expansions we see
that the
above sum equals (note a sign misprint in formula 16.23.10 of \cite{19a})
\begin{equation}
4 q\int_{b/\K}^1\dd z\frac{\dd}{\dd q}\frac{\K}{\sn\,(z\K)}=\frac{\K^2\Z}{2\pi^2\sn\,b},
\label{eq:22}
\end{equation}
with Jacobi's Zeta function $\Z=\Z (b,\kk)$ (in the following we suppress the arguments if they are $b$, $\kk$).
We have finally arrived at
\begin{eqnarray}
\langle \tilde{S} \rangle & = & \int_0^{2\K}\dd\xi\, S(\xi)\; =\; \Z + \frac{{\rm cn}\, b \,
{\rm dn}\, b}{{\rm sn}\, b},
\nonumber \\
\langle \tilde{S}^2 \rangle & = & \int_0^{2\K}\dd\xi\, S(\xi)^2\; =\; \frac{1}{{\rm sn}^2b}+1-\kk^2 - 2 \frac{\bf E}{\K}.
\label{eq:23}
\end{eqnarray}

For later convenience we introduce three basic functions of $b$ and $\kk$,
\begin{equation}
s=\frac{1}{{\rm sn}^2 b},  \qquad t=\frac{\cn\, b\, \dn\, b}{\sn^{3} b}, \qquad u=1-\EK  ,
\label{eq:24}
\end{equation}
where $s$ and $t$ are related by the elliptic equation
\begin{equation}
t^2= s(s-1)(s-\kk^2).
\label{eq:25}
\end{equation}
The spatial averages can then be written in the compact notation
\begin{eqnarray}
\label{eq:26}
\langle \tilde{S}\rangle & = & \Z + t/s , \nonumber \\
\langle \tilde{S}^2\rangle  & = & s-1-\kk^2+2u.
\end{eqnarray}
Averages over higher powers or derivatives of $\tilde{S}$ can also be evaluated in closed form, see Eq.~(\ref{eq:91}).

\vskip 0.1cm
{\em 7) Physical origin}
\vskip 0.1cm
We close this section with a short digression on the role this form of the scalar potential plays in condensed matter physics
and how it was deduced in this context. We first became aware of this ansatz through works on the bipolaron lattice for
non-degenerate conducting polymers \cite{18a}, \cite{21}, \cite{18}--\cite{22}. The following discussion is based on
these references as well as on mathematical literature on periodic solitons \cite{23,24}.

Let us first recall the results for single baryons in the massive GN model \cite{9,9a} (or, equivalently, polarons and
bipolarons in quasi-one-dimensional condensed matter systems \cite{8}). The scalar potential for a baryon has the form
\begin{equation}
S(x) =  1+y \left[ \tanh(yx-c_0)-\tanh (yx+c_0)\right],
\qquad
c_0 = \frac{1}{2} \,{\rm artanh}\, y.
\label{eq:27}
\end{equation}
The parameter $y$ depends on the bare fermion mass and the number of valence fermions. Here
the corresponding isospectral potentials $U_{\pm}$ in the second order equation (\ref{eq:4}) are given
by the simplest P\"oschl-Teller potential \cite{26}
\begin{equation}
S^2 \pm S' = - \frac{2 y^2}{\cosh^2 (yx \pm c_0)}
\label{eq:28}
\end{equation}
and differ only by a translation in space (they are ``self-isospectral" in the terminology of Ref.~\cite{27}).
Their distinguished feature is the fact that they are reflectionless; indeed, they are the unique reflectionless
potential with a single bound state. It is well known that static solutions of the GN model must correspond to
reflectionless Schr\"odinger potentials \cite{7,28}. The fact that the ansatz (\ref{eq:27}) leads to self-consistency
as shown in Refs.~\cite{9,9a} is therefore quite plausible.

Take now a lattice of infinitely many, equidistant P\"oschl-Teller potential wells. As discussed in Refs.~\cite{21,20},
the lattice sum can be performed yielding a Lam\'e-type potential,
\begin{equation}
\sum_{n=-\infty}^{\infty} \frac{1}{\cosh^2(x-nd)} = \left( \frac{2\kk {\K'}}{\pi}\right)^2 \left[\frac{\bf E'}{\kk^2{\K'}}
- {\rm sn}^2\left( \frac{2{\K'}}{\pi} x \right) \right].
\label{eq:29}
\end{equation}
Comparing the spatial period of both sides of Eq.~(\ref{eq:29}), we can relate $d$ and $\kk$ via
\begin{equation}
d = \pi \frac{\K}{\K'}.
\label{eq:30}
\end{equation}
How does the fact that the single potential wells (\ref{eq:28}) are reflectionless manifest itself in the periodic extension
(\ref{eq:29})?
This has been discussed in mathematical physics \cite{23} and condensed matter physics \cite{29} some time ago:
The periodic potential has a single gap (or, in general, a finite number of gaps), in contrast to
generic periodic potentials with infinitely many gaps. Thus reflectionless potentials generalize to
``finite band potentials" as one proceeds from a single well to a periodic array.
In the same way as the sech$^2$-potential is the unique reflectionless potential with one bound state,
the ${\rm sn}^2$-potential is the unique single band potential.
Guided by these considerations, let us try to find the most general superpotential of the Lam\'e potential (plus
an additive constant).
Allowing for a translation in space between $U_{\pm}$ and using an appropriately rescaled coordinate
$\xi$, we have to solve the equations
\begin{eqnarray}
\tilde{U}_+ \ = \ \tilde{S}^2 + \tilde{S}' &=& 2 \kk^2\sn^2(\xi+b) + \eta,
\nonumber \\
\tilde{U}_- \ = \ \tilde{S}^2 - \tilde{S}' &=& 2 \kk^2\sn^2 \xi + \eta
\label{eq:31}
\end{eqnarray}
for $\tilde{S}$ or, equivalently, [compare Eqs.\ (\ref{eq:10}, \ref{eq:16})]
\begin{eqnarray}
\tilde{S}^2 &=& \kk^2 [\sn^2 (\xi+b) + \sn^2 \xi] + \eta,
\nonumber \\
\tilde{S}' &=& \kk^2 [\sn^2 (\xi+b) - \sn^2 \xi].
\label{eq:32}
\end{eqnarray}
Differentiating the upper equation using
\begin{equation}
(\sn\,\xi)' =  \cn\, \xi \, \dn\, \xi
\label{eq:33}
\end{equation}
and dividing the result by the lower equation yields exactly the ansatz (\ref{eq:5}) in the equivalent form (\ref{eq:14}).
As a by-product, by specializing Eq.~(\ref{eq:32}) to $\xi=0$ we can determine the constant $\eta$ [cf.\ Eq.\ (\ref{eq:20})],
\begin{equation}
\eta = \frac{1}{{\rm sn}^2 b}-1-\kk^2.
\label{eq:34}
\end{equation}
The scale factor $A$ in Eq.~(\ref{eq:5}) is not constrained
by these considerations. Thus we conclude that the ansatz (\ref{eq:5}) is the most general Dirac potential
leading to a single gap Lam\'e potential (plus constant) in the corresponding second order equations.
This makes it a good starting point for finding periodic, static solutions.

It could be shown in \cite{12} that the ansatz (\ref{eq:14}) describes the zero temperature limit of the massive GN model.
Regarding $S(x)$ the only novelty is now that all three parameters $A$, $\kk$, $b$ will be varied independently whereas at
$T=0$ the parameter $A$ was given by the density and only $\kk$, $b$ were to be determined by a minimizing
procedure. In the chiral limit $m_0=0$ we already found in \cite{16} that the form of the scalar potential at $T=0$ extends
to finite
temperature without any other change. This result encouraged us to follow a similar strategy in the massive case.

The only way of further generalizing the ansatz would be to go to higher genus Lam\'e potentials, increasing the number
of bands, but for physical reasons we do not expect this to be necessary to describe the thermodynamics of the
GN model. Notice that the simple and intuitive relation between the single baryon and the crystal exhibited in
Eq.~(\ref{eq:29}) is hidden in the corresponding Dirac potentials (\ref{eq:27}) and (\ref{eq:5}) due to the
non-linear relationship between $S$ and $U_{\pm}$.

\section{Minimization of the grand canonical potential and proof of self-consistency}

In this section, we employ the ansatz (\ref{eq:5}) for the scalar potential $S(x)$, minimize the grand
potential with respect to the three parameters $A$, $b$, $\kk$ and prove the self-consistency of the scalar potential.

Collecting the information derived in the previous section we have transformed the Schr\"o\-dinger type eigenvalue
equation (\ref{eq:4})
into the single gap Lam\'e equation
\begin{equation}
\left(-\frac{\partial^2}{\partial \xi^2} + 2\kk^2 \sn^2\xi\right)\phi_+={\cal E} \phi_+,
\label{eq:35}
\end{equation}
and a shifted ($\xi\to\xi +b$) equation for $\phi_-$. The Lam\'e eigenvalues $\mathcal{E}$ are related to the Dirac
energies $E$ through
\begin{equation}
\label{eq:36}
{\cal E}=\frac{E^2}{A^2}-\frac{1}{\sn^2 b}+1+\kk^2 .
\end{equation}
The solutions of  Eq.~(\ref{eq:35}) are well known, see e.g.~Refs.~\cite{30,31}.
For logical completeness we will sketch the results to the extent needed and follow the method used in the chiral
limit \cite{16} to transform the grand potential and the thermal expectation value of $\bar{\psi}\psi$ into complex
contour integrals. Already in the previous section it turned out to be advantageous to extend functions into the
complex plane and use results from complex function theory. Whereas in \cite{16} such an extension was used as a
welcome simplification, here this step is vital to solve a much more intricate problem. It is remarkable
how well the physically real problem fits into the language of complex analysis.
\newpage
\vskip 0.1cm
{\em 1) Eigenfunctions and eigenvalues}
\vskip 0.1cm
We review the properties of the normalized (dimensionless) eigenfunctions of Eq.~(\ref{eq:35}) which by abuse
of notation are again denoted by $\psi=(\phi_+,\phi_-)^T$.

It is possible to give a representation of $\phi_+$ in terms of H,
$\Theta$, and $\Z$, the Jacobi eta, theta, and zeta function \cite{16},
\begin{equation}
\phi_+(\xi)={\cal N}\frac{{\rm H}(\xi+\alpha)}{\Theta(\xi)}{\rm e}^{-\Z(\alpha)\xi},\quad
|{\cal N}|^2=\frac{\K\kk\kk'}{\pi\Theta^2(\alpha)|\dn^2\alpha-\E/\K|}.
\label{eq:37}
\end{equation}
To any Lam\'e-eigenvalue $\cal E$ there exist two complex conjugate eigenfunctions parameterized by $\pm\alpha$.
For $\pm\alpha=\K+\ii\K'\,...\,\K$ one obtains the solutions of the lower band $\kk^2\leq{\cal E}\leq 1$ whereas
$\pm\alpha=0\,...\,\ii\K'$ provides the upper band ${\cal E}\geq 1+\kk^2$.

The eigenfunctions are presented in the form anti-periodic function times complex phase. This gives them a definite
Bloch momentum which is opposite for $\alpha$ and $-\alpha$. The normalization $\cal N$ is chosen such that the
spatial average obeys $\langle|\phi_+|^2\rangle=1/2$ [see Eq.\ (\ref{eq:41})]. The energy eigenvalue together with
the Bloch momentum give a parametric representation of the dispersion relation,
\begin{equation}
{\cal E}=1+\kk^2\cn^2\alpha,\quad p=-\ii\Z(\alpha)\pm\frac{\pi}{2\K},
\label{eq:38}
\end{equation}
where we had to add the offset $\pm\pi/(2\K)$ to the momentum to give the periodic lowest energy solution
$\alpha=\pm(\K+\ii\K')$ zero Bloch momentum. From the dispersion relation we can calculate the density of states and
determine its high momentum asymptotics,
\begin{eqnarray}
&&\frac{\dd p}{\dd{\cal E}}=\frac{|{\cal E}-\kk^2-\E/\K|}{2\sqrt{({\cal E}-\kk^2)({\cal E}-1)({\cal E}-1-\kk^2)}},\nonumber\\
\label{eq:38a}
&&{\cal E}=p^2+2(1-\E/\K)+{\cal O}(p^{-2}).
\end{eqnarray}
It is worthwhile noting that $\tilde{S}(\xi)=\left.-\partial_\xi\ln\phi_+\right|_{\alpha=-b-\xi}$.

With $\phi_+$ given, $\phi_-$ follows from the Dirac equation (\ref{eq:2}),
\begin{equation}
\phi_-=-\frac{1}{E}\left(\frac{\partial}{\partial x}+S\right)\phi_+.
\label{eq:39}
\end{equation}
From this equation we see that $\phi_-$ has the same Bloch momentum as $\phi_+$ (as physically demanded).
Moreover we know that $\phi_-$ solves the $b$-shifted Lam\'e equation. This makes $\phi_-(\xi)$ a multiple of
$\phi_+(\xi+b)$. To determine the ratio we observe that the reflected
and charge conjugated Dirac spinor $\hat\psi(\xi)=-\gamma_0\psi(-b-\xi)^\ast=(\phi_-^\ast(-b-\xi),\phi_+^\ast(-b-\xi))^T$
fulfills the Dirac equation (\ref{eq:2}) for the same energy eigenvalue as $\psi$ and is thus a linear combination of
$\psi$ and $\psi^\ast$. Because $\phi_\pm(-b-\xi)^\ast$ have the same Bloch momentum as $\phi_\pm(\xi)$ we conclude
that $\hat\psi$ is in fact proportional to $\psi$. In components this means $\phi_-^\ast(-b-\xi)=c\phi_+(\xi)$ and $\phi_+^\ast(-b-\xi)
=c\phi_-(\xi)$. Upon iteration we find $cc^\ast=1$. From the explicit form of $\phi_+$, Eq.~(\ref{eq:37}), we know that
$\phi_+(-\xi)^\ast=\pm\phi_+(\xi)$, depending on the energy band. We conclude that $\phi_-(\xi)$ differs from
$\phi_+(\xi+b)$ by a complex phase,
\begin{equation}
\phi_-(\xi)=\ee^{\ii\varphi}\phi_+(\xi+b),\quad\hbox{with } \varphi\in {\rm I} \! {\rm R}.
\label{eq:40}
\end{equation}
This makes the spatial average of their squared absolute values equal and we obtain
\begin{equation}
\langle \psi^{\dagger} \psi\rangle=\langle|\phi_+|^2\rangle+\langle|\phi_-|^2\rangle=2\langle|\phi_+|^2\rangle=1
\label{eq:41}
\end{equation}
as demanded.

Finally we calculate $\bar\psi\psi(\xi)$ and $\psi^\dagger\psi(\xi)$, the scalar and baryon densities of a single orbit.
We find \cite{16}
\begin{equation}
\bar\psi\psi=\frac{1}{E}(\partial_x+2S)|\phi_+|^2,\quad |\phi_+|^2=\frac{1}{2}\frac{{\cal E}-\kk^2-\dn^2\xi}{{\cal E}-\kk^2-\E/\K}.
\label{eq:42}
\end{equation}
We express $\cal E$ in terms of $E$ (Eq.~\ref{eq:36}), use the identity
\begin{equation}
\frac{1}{\sn^2 b}-1+\dn^2\xi=\frac{\sn\,b\,\cn\,\xi\,\dn\,\xi+\sn\,\xi\,\cn\,b\,\dn\,b}{\sn^2 b\,\sn\,(b+\xi)}
\label{eq:43}
\end{equation}
and Eqs.~(\ref{eq:24}) to arrive at the result
\begin{equation}
\bar{\psi}\psi = \frac{(E/A) \tilde{S}-t/(E/A)}{(E/A)^2-s+u}  .
\label{eq:44}
\end{equation}
From (\ref{eq:40}) we know that $\psi^\dagger\psi(\xi)=|\phi_+(\xi)|^2+|\phi_+(\xi+b)|^2$. With Eqs.\ (\ref{eq:16}) and
(\ref{eq:42}) this yields
\begin{equation}
\psi^\dagger\psi = \frac{(E/A)^2+(\tilde{S}^2-3s+1+\kk^2)/2}{(E/A)^2-s+u}.
\label{eq:45}
\end{equation}

\vskip 0.1cm
{\em 2) Grand canonical potential, scalar condensate, and baryon density}
\vskip 0.1cm
In Hartree-Fock approximation, the grand canonical potential density per flavor consists of the independent particle
contribution $\Psi_1$ and the double counting correction $\Psi_2$,
\begin{eqnarray}
\Psi\; & = & \Psi_1 + \Psi_2 ,
\nonumber \\
\Psi_1 &=& -\frac{1}{\beta\pi} \int_0^{\Lambda/2}\!\!\!\dd q \, \ln \left[\left(1+\ee^{-\beta(E-\mu)}\right)\left(1+\ee^{\beta (E+\mu)}
\right)\right], \nonumber \\
\Psi_2 & = & \frac{1}{2Ng^2 \ell}\int_0^\ell\!\!\dd x (S(x)-m_0)^2  .
\label{eq:46}
\end{eqnarray}
Here, $\Lambda/2$ is the UV cutoff, $\ell=2{\K}/A$ denotes the spatial period of $S(x)$ and $q$ is the
Bloch momentum. The fact that $m_0 \neq 0$ manifests itself in two places. Via $S(x)$ it influences the single particle
energies $E$ in $\Psi_1$ and it leads to the shift $S(x) \to S(x)-m_0$ in $\Psi_2$. Let us first transform the momentum
integral in $\Psi_1$ into an integral over Dirac energies. Scaling out the factor $A$ via
\begin{equation}
p=\frac{q}{A}, \qquad \w=\frac{E}{A}
\label{3.1a}
\end{equation}
and using Eqs.~(\ref{eq:24}) the change of integration measure and cutoff is given by [see Eqs.\ (\ref{eq:36}, \ref{eq:38a})],
\begin{eqnarray}
\frac{\dd p}{\dd \w} &=& \frac{\w (\w^2- s+u)}{\pm\sqrt{W}} ,\quad W=(\w^2- s+1)(\w^2- s+\kk^2)(\w^2- s),
\nonumber \\
\Lambda_\w &=& \w(\Lambda/2)\;=\;
\frac{\Lambda}{2A} +\frac{A\langle \tilde{S}^2 \rangle}{\Lambda} + {\rm O}(\Lambda^{-3}) .
\label{eq:47}
\end{eqnarray}
The plus sign in front of $\sqrt{W}$ refers to the upper band, the minus sign to the lower band.
Due to the quadratic divergence of the integral, it is necessary to expand $\Lambda_\w$ to order $\Lambda^{-1}$.
With the shorthand notation
\begin{equation}
a=\beta A,\quad \nu=\beta\mu
\label{eq:48}
\end{equation}
$\Psi_1$ can be written as the following integral over the energy bands $\sqrt{s-1}\leq\w\leq\sqrt{s-\kk^2}$ and
$\w\geq\sqrt{s}$,
\begin{equation}
\pi\beta^2  \Psi_1  = -a\left( \int_{\sqrt{s-1}}^{\sqrt{s-\kk^2}}\!\!\dd\w+\int_{\sqrt{s}}^{\Lambda_\w}\!\!\dd\w\right)
\frac{\dd p}{\dd \w }
\ln\left[(1+\ee^{-a\w+\nu})(1+\ee^{a\w+\nu})\right] .
\label{eq:49}
\end{equation}
Following \cite{16}, we combine the integral over both energy bands (including the sign change in $\sqrt{W}$)
as well as over positive and negative energy modes to the real part of a line integral in the complex $\w$-plane. The
path of integration runs above (or below) the real axis and the sheet of $\sqrt{W}$ is chosen such that
$\sqrt{W}|_{\w=+\infty}=+\infty$ (see Fig.~1) yielding
\begin{equation}
\pi\beta^2  \Psi_1  = -a \re\lim_{\e\to 0}\int_{-\Lambda_\w +\ii \e }^{\infty+\ii \e} \dd\w
\frac{\w(\w^2- s+ u)}{\sqrt{W}} \ln\left(1+\ee^{-a\w +\nu}\right) .
\label{eq:50}
\end{equation}

\vskip 0.2cm
\begin{figure}[ht]
\begin{center}
\epsfig{file=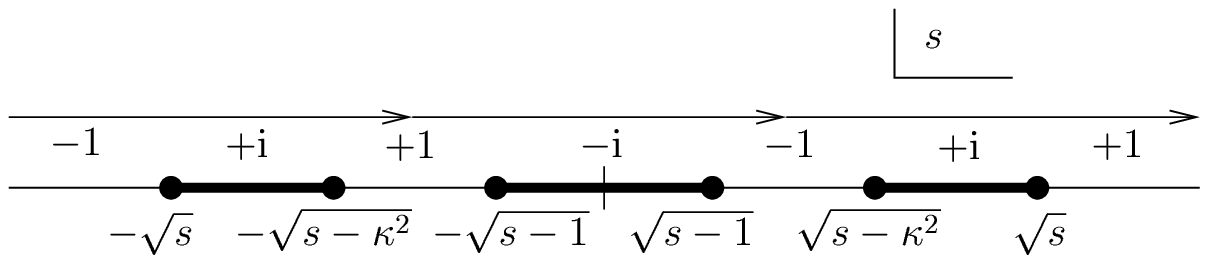,width=12.0cm}
\vskip 0.2cm
\caption{Path of integration in the complex $\w$-plane
running slightly above the real axis. The fat lines indicate branch cuts of $\sqrt{W}$ in Eqs.\ (\ref{eq:47}, \ref{eq:50})
corresponding to the gaps in the Dirac-Hartree-Fock equation (\ref{eq:2}).
We have indicated the complex phase of $\sqrt{W}$ above the real line. Note that $\sqrt{W}$ is antisymmetric under reflections
about the imaginary axis.}
\end{center}
\end{figure}
\vskip 0.2cm

The contribution of $\Psi_2$ was calculated in (\ref{eq:26}),
\begin{eqnarray}
\Psi_2 &=& \frac{1}{2Ng^2} \left( \langle S^2 \rangle - 2 m_0 \langle S \rangle+m_0^2 \right)\nonumber\\
&=& \frac{1}{2Ng^2\beta^2} \left(a^2(s-1-\kk^2+2u) - 2a\beta m_0(\Z + t/s) + \beta^2m_0^2\right).
\label{eq:51}
\end{eqnarray}
With the technique used to rewrite $\Psi$ we transform the thermal expectation values of the scalar potential and the baryon density
 into complex integrals.
Using Eqs.\ (\ref{eq:44}) and (\ref{eq:47}) we obtain
\begin{eqnarray}
\langle \bar{\psi}\psi\rangle_{\mathrm{th}} &=  & \frac{1}{\pi} \int_0^{\Lambda/2}\!\!\dd q\,\bar{\psi}\psi \left(\frac{1}
{\ee^{\beta(E-\mu)}
+1}-\frac{1}{\ee^{-\beta(E+\mu)}+1}\right) \nonumber \\
& = & \frac{a}{\pi\beta} \re \lim_{\e\to 0}\int_{-\frac{\Lambda\beta}{2a}+\ii\e}^{\infty+\ii\e} \!\!\dd \w\,
\frac{\w^2\tilde{S}
-t}{\sqrt{W}}\frac{1}{\ee^{a\w-\nu}+1},
\label{eq:52}
\end{eqnarray}
and analogously with Eq.\ (\ref{eq:45}),
\begin{equation}
\rho(x)=\langle\psi^\dagger\psi\rangle_{\mathrm{th}}=\frac{a}{\pi\beta} \re \lim_{\e\to 0}\int_{-\frac{\Lambda\beta}{2a}
+\ii\e}^{\infty+\ii\e} \!\!\dd \w\,
\frac{\w[\w^2+(\tilde{S}^2-3s+1+\kk^2)/2]}{\sqrt{W}}\frac{1}{\ee^{a\w-\nu}+1}.
\label{eq:53}
\end{equation}
Because the integrals in (\ref{eq:52}) and (\ref{eq:53}) are only logarithmically divergent it is sufficient to keep the leading term
of $\Lambda_\w$. With Eqs.\ (\ref{eq:50}) and (\ref{eq:51}) we confirm that the spatial average of the baryon density
satisfies $\langle\rho(x)\rangle = - \partial \Psi/\partial \mu $.

\vskip 0.1cm
{\em 3) Self-consistency}
\vskip 0.1cm
We have to minimize $\Psi$ with respect to $A,b,\kk$ for fixed $\mu,\beta,\gamma$ and prove that at the minimum the
self-consistency condition
\begin{equation}
\langle \bar{\psi}\psi\rangle_{\mathrm{th}} = -\frac{1}{Ng^2} (S(x)-m_0)
\label{eq:54}
\end{equation}
holds.

It is convenient to rescale the integration variable in (\ref{eq:50}) by $\w\to\w /a$ before differentiation.
The contour of integration
stays off
the singularities of the integrand by a minimum distance of $\e$ (the limit $\e\to 0$ is only needed at the far
left end of the integration contour). This allows us to interchange differentiation and integration
and we obtain equations with integrands of the generic form (polynomial in $\w)\times W^{-3/2}\ln[1+\exp(-a\w+\nu)]$.
Noting that $\partial_\w\ln[1+\exp(-a\w+\nu)]=-a/[\exp(a\w -\nu)+1]$ we see that we gain an equally looking expression
upon applying
a partial integration on the scalar condensate given by Eq.\ (\ref{eq:52}). Comparing coefficients of the polynomials such
that the self-consistency condition (\ref{eq:54}) holds on the level of integrands yields an over-determined system of
equations which
happens to have a solution. This solution actually gives the full self-consistency condition.

In order to present this solution in a most concise way we use the freedom to minimize $\Psi$ with respect
to any set of
independent variables. Specifically, we propose to replace $A$, $b$, $\kk$ by the spatial averages
$\langle S \rangle$, $\langle S^2 \rangle$ and the spatial period $\ell$ of $S$. The stationarity conditions for the
grand potential then read
\begin{eqnarray}
\frac{\partial}{\partial \langle S \rangle} \pi\beta^2\Psi & = &  \beta F_0 \ =\ 0,
\nonumber \\
\frac{\partial}{\partial \langle S^2 \rangle}\pi\beta^2\Psi & = &  \frac{\beta^2}{2} F_1 \ =\ 0,
\nonumber \\
\frac{\partial}{\partial \ell} \pi\beta^2\Psi & = & \frac{a^2}{\ell} F_2 \ = \ 0,
\label{eq:55}
\end{eqnarray}
where we have introduced functions $F_i$ which play a key role in the following. The factors in front of the $F_i$ have been
included for later convenience.
In the limit $m_0\to0$ we find that $F_1$ and $F_2$ equal the corresponding functions in \cite{16}.
The derivatives with respect to the new, composite variables can be taken trivially in the case of $\Psi_2$,
\begin{equation}
\frac{\partial\Psi_2}{\partial\langle S\rangle}=-\frac{m_0}{Ng^2},\quad
\frac{\partial\Psi_2}{\partial\langle S^2\rangle}=\frac{1}{2Ng^2},\quad
\frac{\partial\Psi_2}{\partial\ell}=0.
\label{eq:56}
\end{equation}
Unfortunately this is not true for $\Psi_1$ available only in terms of the original variables $A$, $b$, $\kk$ in Eq.~(\ref{eq:50}).
In order to compute $F_i$, we therefore invoke the chain rule,
\begin{equation}
\left(\begin{array}{c}
\partial_b \\ \partial_A \\ \partial_{\kk}
\end{array} \right) \pi\beta^2\Psi =
\left(\begin{array}{ccc}
\frac{\partial \langle S\rangle}{\partial b} & \frac{\partial \langle S^2\rangle}{\partial b} & \frac{\partial \ell}{\partial b} \\
\frac{\partial \langle S\rangle}{\partial A} & \frac{\partial \langle S^2\rangle}{\partial A} & \frac{\partial \ell}{\partial A} \\
\frac{\partial \langle S\rangle}{\partial \kk} & \frac{\partial \langle S^2\rangle}{\partial \kk} & \frac{\partial \ell}
{\partial \kk}
\end{array}\right)
\left(\begin{array}{c}
\beta F_0\\ \beta^2 F_1/2\\ a^2F_2/\ell
\end{array}\right) .
\label{eq:57}
\end{equation}
Upon inverting the Jacobian matrix on the right hand side, the $F_i$ can be expressed
in terms of  more accessible derivatives of $\Psi$.
Details are given in Appendix A. We arrive at
\begin{eqnarray}
F_0 &=&  -\frac{\pi\beta m_0}{Ng^2} - t\re\lim_{\e\to 0}\int_{-\infty+\ii\e}^{\infty+\ii\e}
 \dd \w \left(\frac{\partial}{\partial\w}\frac{1}{\sqrt{W}}\right)\ln\left(1+\ee^{-a\w +\nu}\right)=0, \nonumber \\
F_1 &=& \frac{\pi}{Ng^2}-1+\frac{1}{a}\re \lim_{\e\to 0} \int_{-\frac{\Lambda\beta}{2a}+\ii\e}^{\infty+\ii\e}
 \dd \w \left(\frac{\partial}{\partial\w}\frac{\w^2}{\sqrt{W}}\right)\ln\left(1+\ee^{-a\w +\nu}\right)=0,\nonumber \\
F_2 &=& \frac{1}{a} \re \lim_{\e\to 0}\int_{-\infty+\ii\e}^{\infty+\ii\e} \dd \w
 \Biggl[ \frac{\w (\w^2-s+u)} {\sqrt{W}}\nonumber \\
&&-\;\frac{\partial}{\partial \w} \frac{t \Z - u \w^2 + (\w^2-s+1)(\w^2-s+\kk^2)}{\sqrt{W}} \Biggr]  \ln\left(1+\ee^{-a\w +\nu}\right)=0.
\label{eq:58}
\end{eqnarray}
Note that the integrand in $F_2$ falls off at $\w\to\pm\infty$ fast enough to render the integral convergent. All three
functions $F_i$
vanish at the minimum of the thermodynamic potential (\ref{eq:50}, \ref{eq:51}). After a partial integration we compare $F_0$
and $F_1$ with Eq.\ (\ref{eq:52}) and obtain
\begin{equation}
\langle\bar{\psi}\psi\rangle_{\mathrm{th}}=S(x)\left(\frac{F_1}{\pi}-\frac{1}{Ng^2}\right)+\frac{F_0}{\pi\beta}+\frac{m_0}{Ng^2}.
\label{eq:59}
\end{equation}
At the minimum of the grand potential we obtain self-consistency (\ref{eq:54}) from $F_0=F_1=0$. We do not need the
minimum condition
$F_2=0$ to ensure self-consistency, nor do we need the limit $\ln\Lambda\to\infty$ as long as we drop terms that
are $1/\Lambda$ suppressed.
This latter property becomes significant in condensed matter physics where $\Lambda$ is the finite band width of
the system under consideration.

\section{Phase boundaries and phase diagram}
The main result of this section is the phase diagram of the massive GN model. The diagram is plotted in Fig.\ 3 and
analyzed in
Subsect.\ 4.4. Before that we have to renormalize the model
and solve the minimum equations $F_i=0$. We end this section with an analysis of the phase boundaries and
its singular points.

\vskip 0.1cm
{\em 1) Renormalization}
\vskip 0.1cm
In the version presented so far the massive GN model still contains an UV cutoff $\Lambda$.
For a quantum field theory this state is not acceptable, it is necessary to remove the cutoff and present a renormalized version
of the model. In the wake of this procedure the bare
parameters $Ng^2$, $m_0$, and the cutoff $\Lambda$ are replaced by physical quantities that have to be adjusted
to observables.
Since all UV divergences are due to vacuum effects, there are no new difficulties as compared to the $T=0$ case \cite{12}.
All we need is the vacuum gap equation
\begin{equation}
\frac{1}{Ng^2} = \frac{1}{\pi} (1+ m_0)\ln \Lambda = \frac{1}{\pi} (\ln \Lambda+ \gamma)  .
\label{eq:60}
\end{equation}
Renormalization of $Ng^2$ interacts with the cutoff $\Lambda$ to merely set the scale of the theory. We fix this scale by the
condition that the dynamical fermion mass in the vacuum equals 1. The bare fermion mass $m_0$ however is renormalized to
a physical
parameter $\gamma=m_0\ln\Lambda$ (called ``confinement parameter" in condensed matter physics).
For different values of $\gamma$ one obtains different physical theories, only one of which would be accessible in a
fictitious 1+1-dimensional world.
With $\gamma\geq 0$ we trace through a one parameter family of physical theories where $\gamma=0$
corresponds to
the chiral, massless GN model. Renormalization can be performed by sending the cutoff $\Lambda$ to infinity while carefully
reducing $Ng^2$ and $m_0$ such that the dynamical fermion mass and $m_0\ln\Lambda$ are kept at the values 1 and
$\gamma$, respectively. After renormalization $N$, $g$, $m_0$, $\Lambda$ have disappeared from the theory in favor of
the single parameter $\gamma$.

In the grand potential irrelevant divergent terms $-\Lambda^2/8\pi$ and $-\mu\Lambda/2\pi$ stemming from
the energy and baryon density of the Dirac sea can simply be dropped. A short calculation using the antisymmetry of
$\sqrt{W}$ and
the fact that due to its holomorphy in the upper half plane $\w(\w^2-s+u)/\sqrt{W}-1$ vanishes if integrated above the real
line shows that
an expression for the renormalized grand potential is ($\lambda=\beta \Lambda$)
\begin{eqnarray}
\pi\beta^2\Psi^{\rm ren} &=&  \lim_{\lambda\to\infty} \Biggl[ \frac{\lambda^2}{8} +\frac{a^2 \langle
\tilde{S}^2\rangle}{2}\left(\ln\frac{\lambda}{\beta} +\gamma -1\right) - a\beta\gamma \langle\tilde{S}\rangle \nonumber \\
& &-\; a\,\re \lim_{\e\to 0}\int_{-\frac{\lambda}{2a}+\ii\e}^{\frac{\lambda}{2a}+\ii\e} \dd \w
 \frac{\w(\w^2- s+u)}{\sqrt{W}} \ln \left(2\cosh \frac{a\w -\nu}{2}\right) \Biggr]  .
\label{eq:61}
\end{eqnarray}
Note that we have not only symmetrized the integration domain but also replaced $\Lambda_\w$ by its leading term and
corrected the effect by subtracting
$a^2\langle\tilde{S}^2\rangle/2$. In fact, the right hand side of Eq.\ (\ref{eq:61}) converges with ${\cal O}(\lambda^{-2})$
as $\lambda\to\infty$.

The effect of renormalization on $F_0$ and $F_1$ amounts to using the gap equation (\ref{eq:60}), whereas $F_2$ is
unaffected by renormalization. Due to symmetry (needed for $F_1$) and holomorphy in the upper half plane we may replace
$\ln[1+\exp(-a\w+\nu)]$ by the more symmetric expression $\ln 2\cosh[(a\w-\nu)/2]$. After partially integrating $F_0$ and
$F_1$ we obtain
\begin{eqnarray}
F_0^{\rm ren}&=&-\beta\gamma+atI_0=0,\quad\;\, I_0=\frac{1}{2}\re\lim_{\e\to 0}\int_{-\infty+\ii\e}^{\infty
+\ii\e}
 \dd \w \frac{1}{\sqrt{W}}\tanh\frac{a\w -\nu}{2},\nonumber\\
F_1^{\rm ren}&=&\gamma-\ln\beta-I_1=0,\quad I_1=\lim_{\lambda\to\infty}\left[-\ln \lambda
+\frac{1}{2}\re \lim_{\e\to 0}
\int_{-\frac{\lambda}{2a}+\ii\e}^{\frac{\lambda}{2a}+\ii\e} \dd \w \frac{\w^2}{\sqrt{W}}\tanh
\frac{a\w -\nu}{2}\right],\nonumber\\
F_2&=& \frac{1}{a} \re \lim_{\e\to 0}\int_{-\infty+\ii\e}^{\infty+\ii\e}\!\dd \w
 \Biggl[ \frac{\w (\w^2\!-\!s\!+\!u)} {\sqrt{W}}
-\frac{\partial}{\partial \w} \left(\frac{t \Z\! -\! u \w^2}{\sqrt{W}} + \frac{\sqrt{W}}{\w^2\!-\!s}\right) \Biggr]
\ln\left(2\cosh\frac{a\w\!-\!\nu}{2}\right)\nonumber\\
&=&0.\label{eq:62}
\end{eqnarray}
Note that $I_0$, $I_1$, and $F_2$ have no explicit $\beta$-dependence.

Although we will not make use of it in the
following, let us mention that one can use $F_i=0$ to present $\Psi$ at its minimum in a somewhat simplified way. One obtains
\begin{equation}
\pi\beta^2\Psi^{\rm ren}_{\rm min}=-\frac{\lambda^2}{8}-\frac{a\beta\gamma t}{s} +\frac{a^2}{2}\Biggl[
(s-1-\kk^2)\left(\ln\frac{\lambda}{\beta} +\gamma\right)
+\re\!\!\int\limits_{-\frac{\lambda}{2a}+\ii\e}^{\frac{\lambda}{2a}+\ii\e}
 \frac{\dd\w\sqrt{W}}{\w^2-s}\tanh \frac{a\w -\nu}{2}\Biggr],
\label{eq:63}
\end{equation}
where the limits $\e\to0$ and $\lambda \to \infty$ are understood.

After solving Eqs.~(\ref{eq:62}) we can use Eqs.~(\ref{eq:61}) or (\ref{eq:63}) to calculate the pressure
$P=-\Psi$ as a function of $\mu$, $T$ giving the equation of state. Bulk thermodynamical variables like the (spatially averaged)
baryon density $\rho$, the entropy density $\sigma$ and the energy density $\e$ can then be obtained via
standard thermodynamic relations,
\begin{equation}
\rho =- \frac{\partial}{\partial \mu} \Psi, \qquad \sigma=\beta^2 \frac{\partial}{\partial \beta}\Psi, \qquad \e
=T\sigma  -P+\mu\rho.
\label{eq:64}
\end{equation}

\vskip 0.1cm
{\em 2) Solving the minimum equations}
\vskip 0.1cm
With the physical parameters $\gamma$, $\beta=1/T$, $\nu=\mu/T$ given we have to solve the three equations
$F^{\rm ren}_0=F^{\rm ren}_1=F_2=0$, Eqs.\ (\ref{eq:62}),
simultaneously to obtain the values of the parameters $a$, $b$, $\kk$ at the minimum of the grand potential.
There exists a simple and efficient method to achieve this numerically.

We observe that $F_2$ depends on only four of the six variables involved, namely on $a$, $b$, $\kk$, $\nu$.
For convenience we fix a value of $\kk$ with $0<\kk<1$, of $b$ with $0<b<\K$, and the physical $\nu$.
With $\kk$ and
$b$ given, we calculate $s$, $t$, $u$, $\Z$ and consider the function $F_2(a)$. We find that $F_2(a)=ca^2+{\cal O}(a^4)$
for $a\to0$ [see Eq.\ (\ref{eq:89})]
and that $F_2(a)$ goes to a negative constant for $a\to\infty$. If (here $\psi(n,z)=\partial_z^{n+1}\ln\Gamma(z)$ is the
polygamma function)
\begin{equation}
\nu>\nu_{\rm t},\quad\hbox{with}\quad \re\,\psi\left(2,\frac{1}{2}+\frac{\ii\nu_{\rm t}}{2\pi}\right)=0,
\quad \nu_{\rm t}=1.910668
\label{eq:65}
\end{equation}
we have $c>0$ and the function $F_2(a)$ has a zero at a unique positive value of $a$.
For $\nu<\nu_{\rm t}$ we have $c<0$ and $F_2(a)=0$ has only the trivial solution $a=0$. In the latter case the crystal solution
will not exist and we are in the (simpler) translationally invariant regime studied in the following subsection.
At the moment we restrict ourselves to the case $\nu>\nu_{\rm t}$.

With $F_2(a)$ given as a one-dimensional integral over elementary functions the positive zero will quickly be found.
Now we can calculate the integrals $I_0$ and $I_1$ in Eqs.\ (\ref{eq:62}) which also depend on $a$, $b$, $\kk$,
$\nu$ only. Combining $F_0=0$ and $F_1=0$, we obtain
\begin{equation}
\gamma + \ln \gamma = \ln(atI_0) + I_1,\quad \beta =atI_0/\gamma.
\label{eq:66}
\end{equation}
These equations determine $\gamma$ and $\beta$ for given values of $b$, $\kk$, $\nu$.
Normally, one would like to dial $\gamma$, $\mu$, $T$ and evaluate thermodynamic observables for these values.
Therefore one has to work one's way to the desired values of $\gamma$ and $\beta$ by systematically estimating
parameters $b$ and $\kk$ to start with. If $\gamma$, $\mu$, $T$ lie in the crystal region depicted in Fig.\ 3 the
iteration will terminate quickly.
Only at the phase boundary this procedure becomes less efficient, but there it will be possible to derive alternative
methods by virtue of analytical techniques (see Subsect.\ 4.5).

We close this subsection with the remark that for actually implementing this algorithm on a computer one should transform
the contour integrals into real integrals of smooth functions with compact support. This is always possible as
worked out in detail for a variety of integrals in Appendix B.

\vskip 0.1cm
{\em 3) Homogeneous phase and the ``old" phase diagram}
\vskip 0.1cm
The translationally invariant, homogeneous phase has been studied in \cite{13}. We will briefly review the results and fit them
into the context of this more comprehensive approach.

We saw in Eq.\ (\ref{eq:17}) that in the limits $\kk\to0$ and $\kk\to1$ the scalar potential $S$ tends to a constant.
The single baryon in the case $\kk\to1$ is insignificant in entire space. We see from the
Dirac-Hartree-Fock equation (\ref{eq:2}) that in this case the scalar potential has to be
interpreted as an effective fermion mass $M$. We have
\begin{equation}
S(x)=M=\frac{a\sqrt{s-1}}{\beta}\quad\hbox{for}\quad\kk=0
\quad\quad\hbox{and}\quad\quad M=\frac{a\sqrt{s}}{\beta}\quad\hbox{for}\quad\kk=1.
\label{eq:67}
\end{equation}
The homogeneous phase is accessible in our framework by setting $\kk=0$ or $\kk=1$ and {\em not}
demanding $F_2=0$.
Because the minimum condition $F_2=0$ was not needed for self-consistency we will not lose self-consistency
in the homogeneous phase.
The two bands collapse to a single band and the grand potential in either limit specializes to [after a rescaling
$\w\to\w\beta/a$ and a partial integration in Eq.\ (\ref{eq:61})]
\begin{equation}
\pi\Psi^{\rm ren}_{\rm hom} = \lim_{\Lambda\to\infty} \Biggl[-\frac{\Lambda^2}{8}+\frac{M^2}{2}\left(\ln\Lambda+\gamma\right)
-\gamma M
 +\frac{1}{2}\re\lim_{\e\to 0}\int_{-\frac{\Lambda}{2}+\ii\e}^{\frac{\Lambda}{2}+\ii\e} \!\!\dd \w
 \sqrt{\w^2\!-\!M^2} \tanh\frac{\beta\w -\nu}{2}\Biggr].
 \label{eq:68}
\end{equation}
The limit $\Lambda\to\infty$ can easily be performed. For comparison with other work we give the result in the
original variables,
momentum $q$ and energy $E=\sqrt{q^2+M^2}$,
\begin{equation}
\Psi^{\rm ren}_{\rm hom} = \frac{M^2}{4\pi}(2\ln M-1)+\frac{\gamma M}{2\pi}(M-2)
-\frac{1}{\beta \pi} \int_0^{\infty} {\rm d}q \ln
\left[ \left( 1+ {\rm e}^{-\beta(E-\mu)}\right) \left( 1+ {\rm e}^{-\beta(E+\mu)}\right)\right].
\label{eq:69}
\end{equation}
The minimum condition is obtained by differentiation of $\pi\Psi^{\rm ren}_{\rm hom}$ with respect to $M$.
Comparison with Eqs.\ (\ref{eq:62})
[or Eqs.\ (\ref{eq:73}, \ref{eq:75})] yields
\begin{equation}
0=M\ln M+\gamma(M-1)+M\int_0^{\infty} \frac{\dd q}{E}
\left(\frac{1}{{\rm e}^{\beta(E-\mu)}+1}+\frac{1}{{\rm e}^{\beta(E+\mu)}+1}\right)=\frac{F_0^{\mathrm{ren}}}{\beta}+
MF_1^{\mathrm{ren}}.
\label{eq:70}
\end{equation}
In fact, only $M$, a combination of $a$ and $b$, is of relevance. This reduces self-consistency to a single
linear combination of $F_0$ and $F_1$. A second equation is not needed.
For the moment we study the homogeneous phase in the whole $(\mu,T)$-plane, keeping in mind that the phase 
diagram will change
due to the existence of the crystal phase.

Depending on the parameters ($\gamma,\mu,T$), there may be
one or two local minima with the possibility of a first order phase transition.
Since we could not find a published graph of the phase diagram in ($\gamma,\mu,T$)-space, we
have recalculated it with the result shown in Fig.~2.
\vskip 0.2cm
\begin{figure}[t]
\begin{center}
\epsfig{file=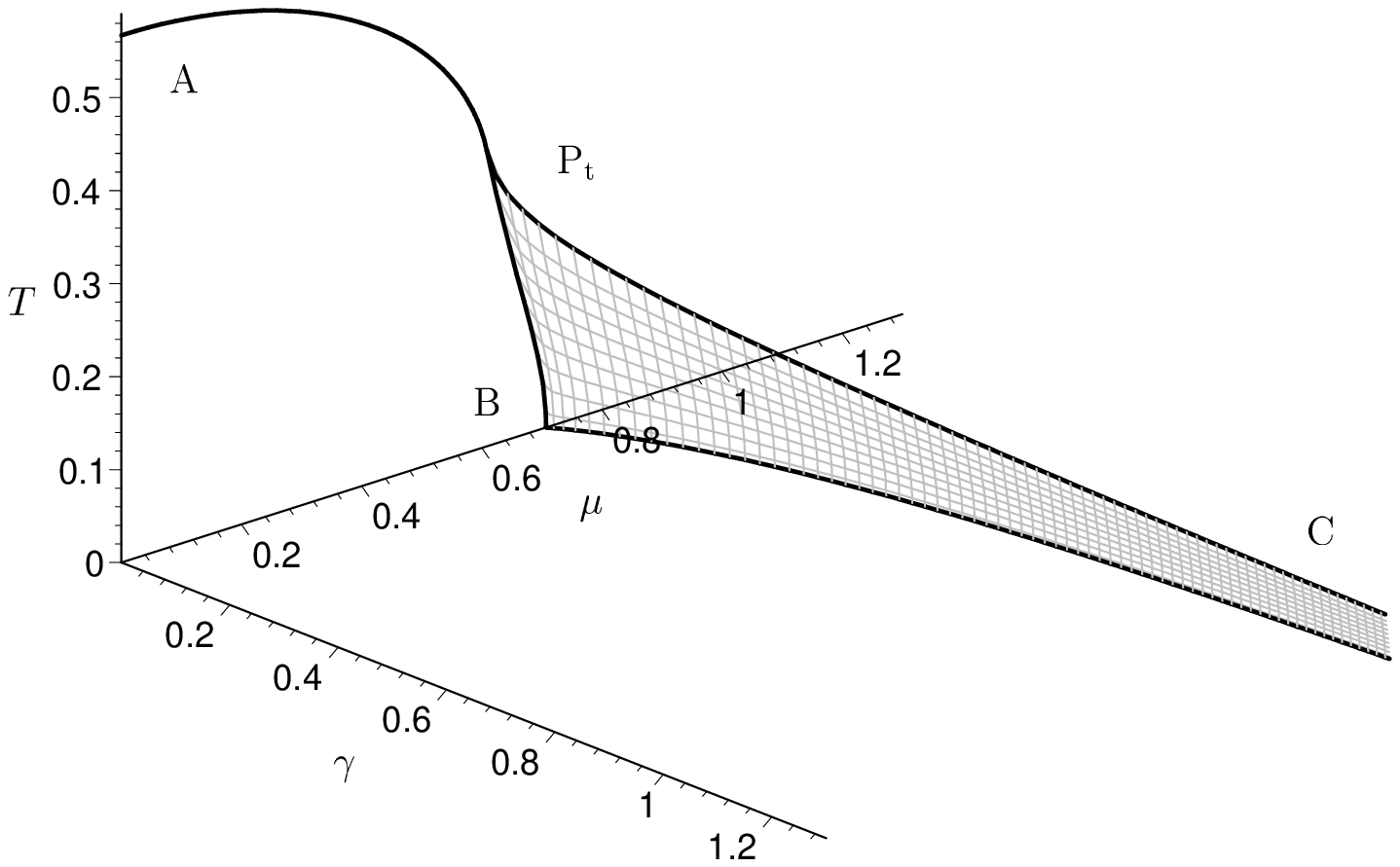,width=9.0cm,angle=270}
\vskip 0.2cm
\caption{``Old" phase diagram of the massive Gross-Neveu model as a function of ($\gamma,\mu,T$), obtained under
the (unjustified) assumption of unbroken translational symmetry. The various phase boundaries are explained
in the main text.}
\end{center}
\end{figure}
\vskip 0.2cm
In the chiral limit ($\gamma=0$), the massive phase
at low $(\mu,T)$ is separated from the chirally restored, massless phase at high $(\mu,T)$ by a critical line AP$_{\rm t}$B
\cite{32}. The upper part  AP$_{\rm t}$ of this line is second order, the lower part P$_{\rm t}$B first order, with a tricritical
point P$_{\rm t}$ separating the two. As the parameter $\gamma$ is switched on,
the second order line disappears in favor of a cross-over where the fermion mass changes rapidly, but smoothly.
The first order line on the other hand survives, ending at a critical point.
The effect of the cross-over shrinks to zero when the critical point is approached.
If plotted against $\gamma$, these
critical points lie on the third curve P$_{\rm t}$C emanating from the tricritical point. For $\gamma>0$
the effective fermion mass $M$ is positive everywhere.
If one crosses the shaded critical ``sheet" in Fig.~2, the mass
changes discontinuously, dropping with increasing chemical potential. More details on the thermodynamic implications
of such
a phase diagram can be found in Ref.~\cite{13}.

\vskip 0.1cm
{\em 4) Crystal phase and the revised phase diagram}
\vskip 0.1cm
According to the discussion of $S(x)$ in Subsect.~2.4, at the phase boundary between homogeneous and crystal
phases, $\kk$
has the value 0 or 1. Only for these limiting values of the elliptic modulus, the spatial modulation of $S(x)$ becomes
insignificant. This observation is the key for deriving the phase boundary equations in the following subsection.
We have computed points of constant $\nu=\mu/T$ varying $b$ from 0 to $\pi/2$ or to $\infty$, respectively.
For $\kk=0$ or 1, varying $\nu>\nu_{\rm t}$ [see Eq.\ (\ref{eq:65})] we obtain $(b,\nu)$-grids which are mapped onto
curved two-dimensional
surfaces in ($\gamma,\mu,T$)-space. These surfaces represent second order phase boundaries (see Subsect.\ 4.5).

\vskip 0.2cm
\begin{figure}[ht]
\begin{center}
\epsfig{file=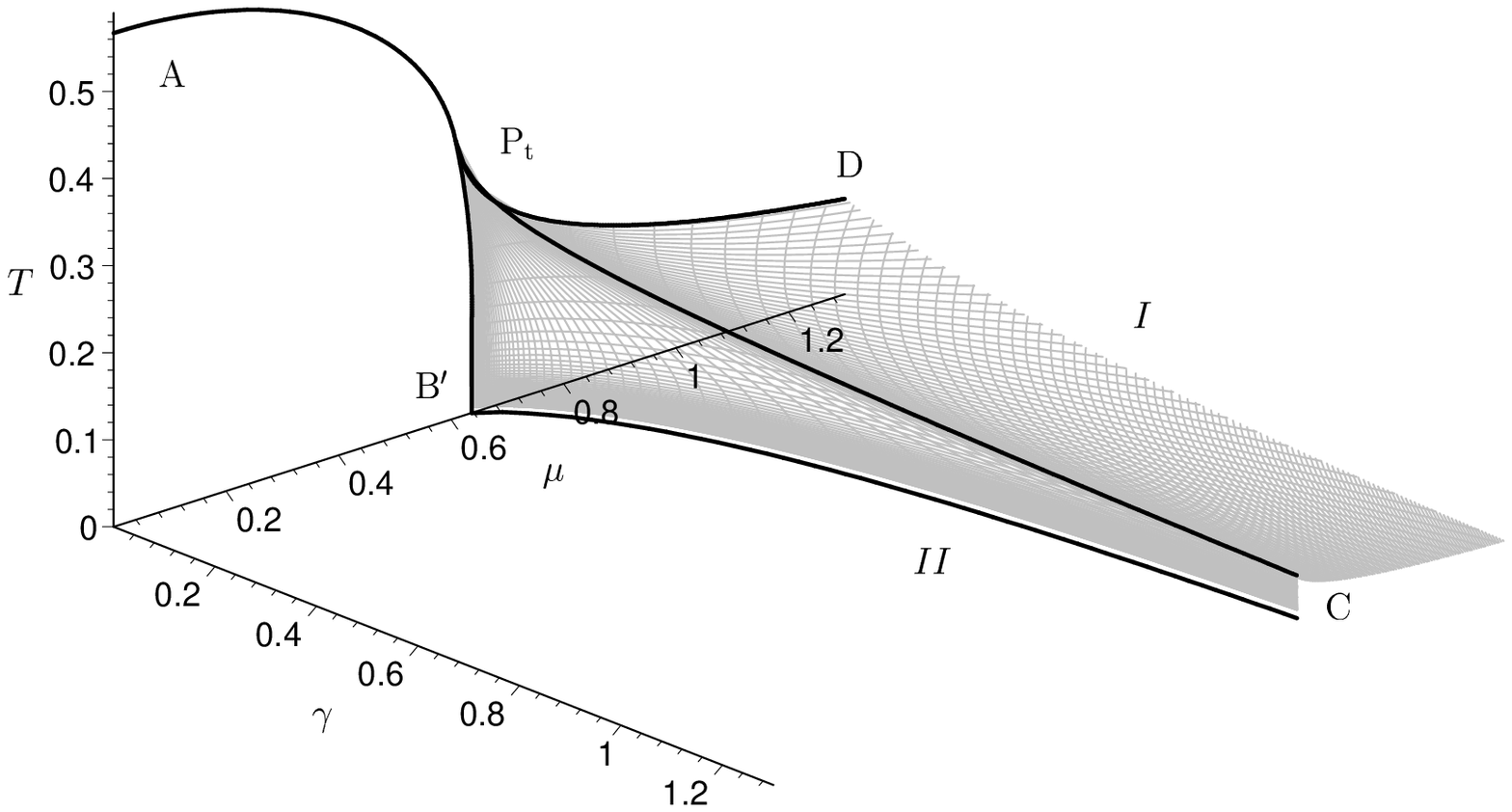,width=9.0cm,angle=270}
\vskip 0.2cm
\caption{Revised phase diagram of the massive Gross-Neveu model. The shaded surfaces $I,II$ separate the kink-antikink
crystal from a massive Fermi gas and correspond to second order phase transitions.}
\end{center}
\end{figure}
\vskip 0.2cm

We now turn to the results, focusing on the phase boundaries as depicted in Fig.~3.
In the $\gamma=0$ plane, the result for the revised phase diagram including the kink-antikink crystal is known
\cite{16}: The second order line AP$_{\rm t}$ of the old phase diagram is unaffected. The first order line
P$_{\rm t}$B of Fig.~2 is replaced by two second order lines P$_{\rm t}$B$'$ and P$_{\rm t}$D delimiting
the kink-antikink crystal phase. The
tricritical point is turned into another kind of multicritical point (called Lifshitz- or Leung point in condensed matter
physics \cite{33,34}), located at the same ($\mu,T$) values. As we turn on $\gamma$, the second order line separating
massive ($M>0$) and massless ($M=0$) phases disappears as a consequence of the explicit breaking of chiral
symmetry. The crystal phase survives at all values of $\gamma$, but is confined
to decreasing temperatures with increasing $\gamma$. For fixed $\gamma$, it is bounded by two second order lines joining in a
cusp. The cusp coincides with the critical point of the old phase diagram but has a significantly different
character. The crystal phase exists and is thermodynamically stable inside the tent-like structure formed out of two sheets
denoted as $I$ and $II$. These sheets are defined
by $\kk=0$ ($I$) and $\kk=1$ ($II$), respectively. The line P$_{\rm t}$C where they join corresponds to $b=0$
and coincides with line P${\rm _t}$C in Fig.~2. The baseline of sheet $II$ in the ($\mu,\gamma$)-plane
has a simple physical interpretation: It reflects the $\gamma$-dependence of the baryon mass in the massive GN
model, or equivalently the critical chemical potential at $T=0$.
The chiral limit $\gamma \to 0$ can be identified with $b \to {\K}$. In this way, one can recover
the simpler 2-parameter ansatz for the self-consistent potential used in \cite{16} from the 3-parameter
ansatz (\ref{eq:5}). Incidentally, we should like to point out that all the fat lines in Fig.~3 representing
the boundaries of the sheets $I,II$ have been taken from sources independent of the present work \cite{13,14,16}.
The fact that the computed surfaces connect very accurately constitutes a test
for our algebraic and numerical calculations.

\vskip 0.2cm
\begin{figure}[ht]
\begin{center}
\epsfig{file=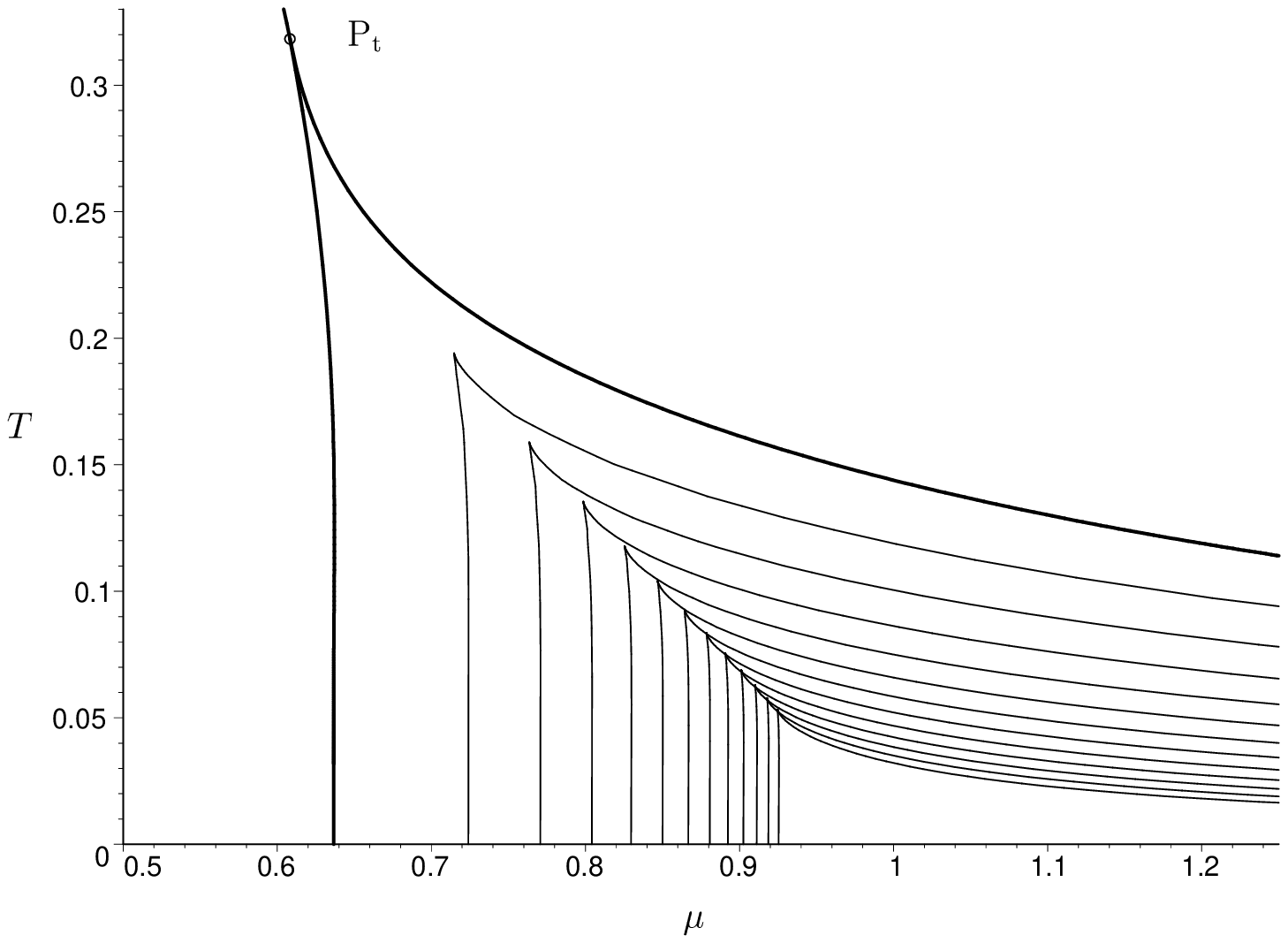,width=9.0cm,angle=270}
\vskip 0.2cm
\caption{Two-dimensional sections $\gamma=$ const.\ through the phase diagram of Fig.\ 3. The fat lines correspond
to $\gamma=0$, the thin lines to $\gamma=0.1...1.2$ (in steps of 0.1), from top to bottom. The position of the cusp agrees with
the critical point in the old phase diagram.}
\end{center}
\end{figure}
\vskip 0.2cm

Another view of the phase diagram is shown in Fig.~4. Here we plot the phase boundaries in the
($\mu,T$)-plane for several values of $\gamma$, i.e., for several bare quark masses. These two-dimensional
graphs correspond to cutting the three-dimensional graph in Fig.~3 by equidistant planes of constant $\gamma$.
They provide a better view of how the two critical lines are joined in the cusp and exhibit that the region in which
the crystal is thermodynamically stable shrinks rapidly with increasing $\gamma$.

\vskip 0.2cm
\begin{figure}[ht]
\begin{center}
\epsfig{file=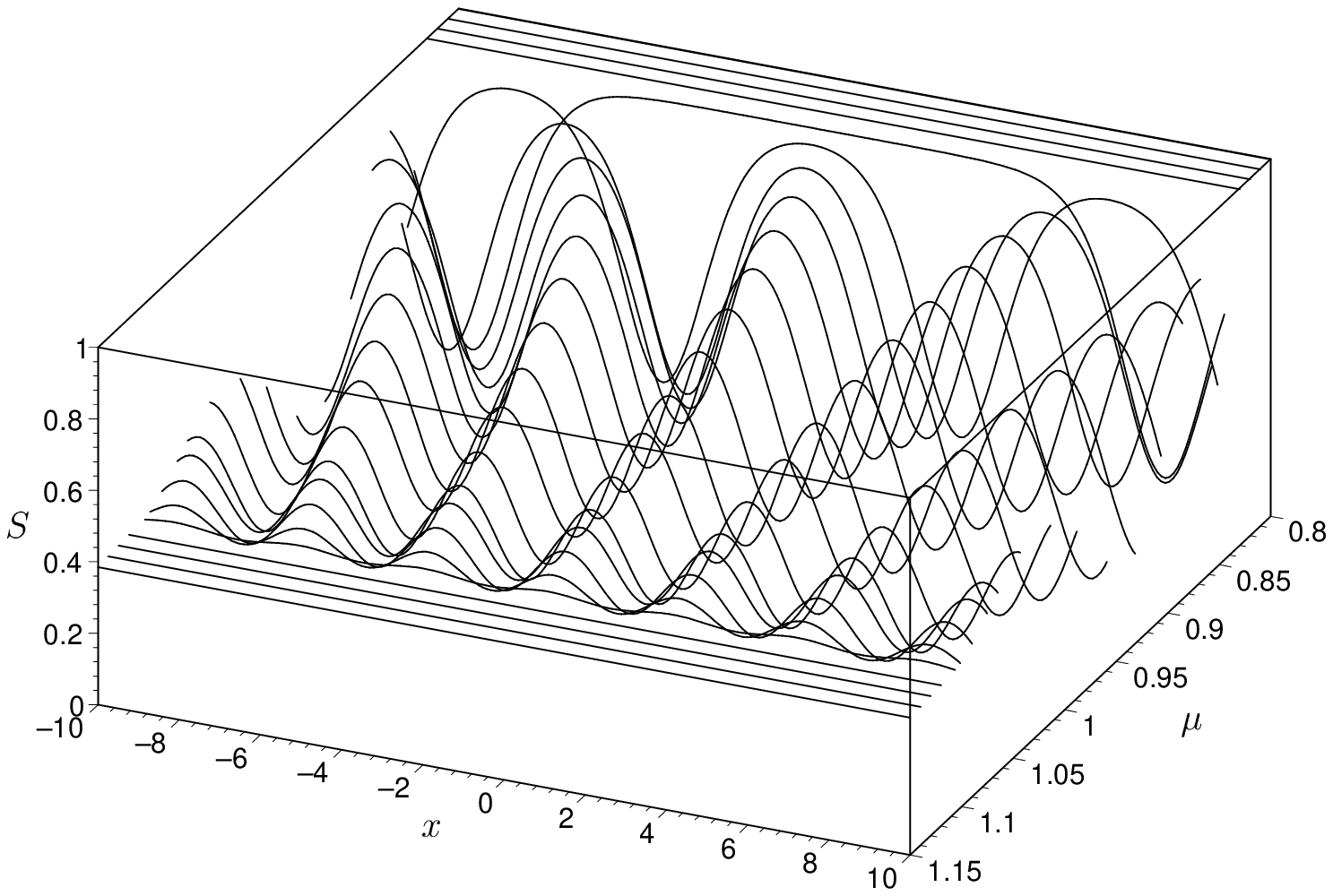,width=9.0cm,angle=270}
\vskip 0.2cm
\caption{Evolution of self-consistent $S(x)$ as one crosses the crystal region on a straight path. As $\mu$ decreases from
1.1 to 0.85, the elliptic modulus $\kk$ varies between 0 and 1. According to the ``old" phase diagram, there would be
a discontinuous jump between two constant values of $S$ at $\mu=0.87$.}
\end{center}
\end{figure}
\vskip 0.2cm

In order to further illustrate the nature of the phase transitions between inhomogeneous and homogeneous
phases, let us remark that outside the tent the scalar potential is $x$-independent, i.e., an effective mass.
If one crosses the tent from side $I$ to side $II$ on some curve, the parameter $\kk$ varies continuously from 0 to 1.
By way of example consider the straight line $\gamma=0.5$, $T=0.05 \mu$ which pierces sheet $I$ at $\mu=1.1$
and sheet $II$ at $\mu=0.85$. Just outside of the crystal phase, the effective mass values we find are
0.40 ($I$) and 1.0 ($II$).
According to the old phase diagram, the mass jumps suddenly from 0.54 to 1.0 upon crossing
the first order sheet at $\mu=0.87$. By contrast, Fig.~5 shows the continuous evolution of $S(x)$ in the exact calculation.
At $\kk=0$, an instability occurs with respect to oscillations of a finite wave number. As $\kk$ increases,
the amplitude of these oscillations become larger whereas the period first grows rather modestly.
When $\kk$ approaches the value 1 the period grows rapidly and, save for distant dents representing widely spaced
baryons, $S(x)$ reaches a constant value
connecting to the translationally invariant phase. At $\kk=1$ the system is instable against single baryon formation.
This subtle interpolation of $S(x)$ between two constants caused by the Peierls instability is only crudely modelled by a
translationally invariant scenario.

Before we examine the phase boundaries and its special points in the next three subsections, let us summarize the
thermodynamics
of the massive GN model in Table 1. It is important to notice that the sheet of first order phase transitions
present in the old phase diagram is entirely superseded by the crystal phase. The grand potential of the GN model
is smooth for any values of $\gamma$ (including the value zero) everywhere in the $(\mu,T)$-plane. Even approaching the
critical points from any direction does not result in a kink in the grand potential:
The effect of the cross-over as well as of crystal formation reduces to zero when the critical point is reached.

\renewcommand{\arraystretch}{1.2}
\begin{table}[ht]
\begin{center}
\begin{tabular}{l|l}
\hline \hline
$\Psi_{\rm hom}$&Eqs.\ (\ref{eq:69}, \ref{eq:70})\\
$\Psi_{\rm cryst}$&Eqs.\ (\ref{eq:61}, \ref{eq:62})\\ \hline
&Surfaces\\\hline
$I$&Eqs.\ (\ref{eq:62}, \ref{eq:73}),\quad 2$^{\rm nd}$ order phase boundary \\
$II$&Eqs.\ (\ref{eq:62}, \ref{eq:75}),\quad 2$^{\rm nd}$ order phase boundary\\\hline
&Lines\\\hline
AP$_{\rm t}$&$\gamma=0,\quad\ln(4\pi T)+\re\,\psi\left(\frac{1}{2}+\frac{\ii\nu}{2\pi}\right)=0$\\
P$_{\rm t}$B$'$&$\gamma=0,\quad\frac{\partial}{\partial a}\frac{1}
{a}\int_0^{\pi/2}
 \frac{\dd\varphi}{\cos\varphi} \im \ln \frac{\Gamma\left(\frac{1}{2}+\frac{\ii}{2\pi}(\nu+a\cos \varphi)\right)}
{\Gamma\left(\frac{1}{2}+\frac{\ii}{2\pi}(\nu-a\cos \varphi)\right)} =0$,\\
&$\ln(4\pi T)+\frac{1}{\pi}\int_0^\pi\dd\varphi \, \re\, \psi\left(\frac{1}{2}+
\frac{\ii}{2\pi}(\nu+a\cos\varphi)\right)=0$\\
P$_{\rm t}$D&$\gamma=0,\quad\ln(4\pi T)+\frac{1}{2}\min\limits_{a\geq 0}\re\left[\psi\left(\frac{1}{2}+\frac{\ii}{2\pi}(\nu+a)\right)
+\psi\left(\frac{1}{2}+\frac{\ii}{2\pi}(\nu-a)\right)\right]=0$\\
B$'$C&$T=0,\quad\theta+\gamma\tan\theta=\pi/2,\quad\sin\theta+\gamma\,\hbox{artanh}(\sin\theta)=\mu\pi/2$\\
P$_{\rm t}$C&Eq.\ (\ref{eq:83})\\\hline
&Points\\\hline
A&$(\gamma,\mu,T)=(0,0,\ee^{\rm C}/\pi)=(0,0,0.566932),\quad {\rm C}=0.577215$ (Euler constant)\\
P$_{\rm t}$&$(\gamma,\mu,T)=(0,\nu_{\rm t}T_{\rm t},T_{\rm t})=(0,0.608221,0.318329)$,\\
&Eq.\ (\ref{eq:65}),\quad $\ln(4\pi T_{\rm t})+\re\,\psi\left(\frac{1}{2}+\frac{\ii\nu_{\rm t}}{2\pi}\right)=0$\\
B$'$&$(\gamma,\mu,T)=(0,2/\pi,0)=(0,0.636619,0)$\\ \hline\hline
\end{tabular}
\renewcommand{\arraystretch}{1}
\caption{Guide through the phase diagram of the massive GN model, using results from \cite{9,12,16}.
In this table $\Gamma(z)$, $\psi(z)$, and $\psi(n,z)$ are the Gamma,
digamma, and polygamma functions and $\nu=\mu/T$.}
\end{center}
\end{table}

\vskip 0.1cm
{\em 5) Phase boundaries}
\vskip 0.1cm
We saw in Eq.\ (\ref{eq:17}) that we lose the crystal shape of $S(x)$ in the limits $\kk\to 0$ and $\kk\to 1$. In the
first case the amplitude of the modulations vanishes, in the second case the frequency of the modulations goes to zero.
Corroborated by numerical results and the $\gamma=0$ limit in \cite{16} we expect to reach the phase boundaries when
$\kk$ tends towards its extremal values 0 and 1 (the boundaries are labelled $I$ and $II$, respectively, in Fig.\ 3).
More precisely, only in the limit $\gamma=0$ we will have two phase
boundaries requiring at least three different phases. For $\gamma>0$ we will find that the only phase boundary between the
crystal and the massive Fermi gas consists of two different sections separated by a distinguished point. This point, the
cusp, will be examined in the next subsection. Here we focus on the behavior of the theory for values of $\kk$
close to 0 or 1.

In the limit $\kk\to0$ the two energy bands combine to a single band $\w^2\geq s-1$. A remnant of the shrunken gap
is a singularity at $\w^2=s$ inside the energy band. It is useful to change variables from $b$ to $s=\sn^{-2}b$
which gives the expansions in $\kk$ a compact form,
\begin{eqnarray}
t\Z&=&\frac{s-1}{2}\kk^2+\frac{(s-1)(s-2)}{16s}\kk^4+\frac{(s-1)(s^2-2)}{32s^2}\kk^6\nonumber\\
&&+\;\frac{(s-1)(41s^3+6s^2-8s-80)}{2048s^3}\kk^8+\mathcal{O}(\kk^{10}),\nonumber\\
u&=&\frac{1}{2}\kk^2+\frac{1}{16}\kk^4+\frac{1}{32}\kk^6+\frac{41}{2048}\kk^8+\mathcal{O}(\kk^{10}).
\label{eq:72}
\end{eqnarray}
Expanding $F_i$ yields [see Eqs.\ (\ref{eq:62})]
\begin{eqnarray}
\left.F_0^{\rm ren}\right|_{\kk=0}&=&-\beta\gamma+as\sqrt{s-1}\left.I_0\right|_{\kk=0}=0, \nonumber \\
\left.F_1^{\rm ren}\right|_{\kk=0}&=&\gamma-\ln\beta-\left.I_1\right|_{\kk=0}=0, \nonumber \\
\lim_{\kk\to0}\frac{16s}{\kk^4}F_2 &=& \re \lim_{\e\to 0}
 \int_{-\infty+\ii\e}^{\infty+\ii\e}\dd\w \frac{\w^2}{\sqrt{\w^2-s+1}(\w^2-s)^2}\tanh\frac{a\w -\nu}{2}=0.
\label{eq:73}
\end{eqnarray}
Solving these equations following the procedure of Subsect.\ 4.2 with $\kk=0$ leads to a relation between the physical
parameters $\gamma$, $\mu$, $T$ which determines phase boundary $I$.

In the limit $\kk\to1$ the procedure is analogous. The lower energy band collapses to a point singularity at $\w^2=s-1$.
We are left with a single band $\w^2\geq s$ with no singularity inside. Expanding in $\kk'=\sqrt{1-\kk^2}$ yields
\begin{eqnarray}
t\Z&=&(s-1)[1-b\sqrt{s}(1-u)]+\frac{2-s+b\sqrt{s}(s-2+u)}{2}\kk'\,^2\nonumber\\
&&+\;\frac{-s(s+1)+b\sqrt{s}(s^2+2s-1-2u)}{16(s-1)}\kk'\,^4+\mathcal{O}(\kk'\,^6),\nonumber\\
u&=&1-\delta-\frac{2-2\delta+\delta^2}{4}\kk'\,^2-\frac{8-12\delta-3\delta^2+8\delta^3}{128}\kk'\,^4
+\mathcal{O}(\kk'\,^6),\quad \delta=\frac{1}{\ln(4/\kk')},\nonumber\\
s&=&\coth^2 b+\left[\frac{b\coth b}{\sinh^2 b}-\coth^2 b\right]\frac{\kk'\,^2}{2}\nonumber\\
&&+\left[\frac{2b^2(3\!+\!2\sinh^2 b)}{\sinh^4 b}-\frac{3b\coth b}{\sinh^2 b}-(3\!-\!2\sinh^2 b)\coth^2 b\right]
\frac{\kk'\,^4}{32}+\mathcal{O}(\kk'\,^6).
\label{eq:74}
\end{eqnarray}
Expanding $F_i$ at $\kk=1$ yields
\begin{eqnarray}
\left.F_0^{\rm ren}\right|_{\kk=1}&=&-\beta\gamma+a\sqrt{s}(s-1)\left.I_0\right|_{\kk=1}=0, \nonumber \\
\left.F_1^{\rm ren}\right|_{\kk=1}&=&\gamma-\ln\beta-\left.I_1\right|_{\kk=1}=0, \nonumber \\
\lim_{\kk\to1}\ln\left(\frac{4}{\kk'}\right)F_2
&=&\frac{1}{a} \re\lim_{\e\to 0}\int_{-\infty+\ii\e}^{\infty+\ii\e}
\dd \w \Biggl[-\frac{\w}{(\w^2-s+1)\sqrt{\w^2-s}}\nonumber\\
&&+\;\frac{\partial}{\partial \w} \frac{b\sqrt{s}(s-1)-\w^2}{(\w^2-s+1)\sqrt{\w^2-s}} \Biggr]
\ln\left(2\cosh\frac{a\w-\nu}{2}\right)=0.
\label{eq:75}
\end{eqnarray}
These equations, too, can be solved with the procedure of Subsect.\ 4.2 yielding the relation that defines phase boundary $II$.
Neither in the case $\kk=0$ nor in the case $\kk=1$ it is possible to formulate the phase boundary condition in
terms of the effective fermion mass $M$ [Eq.\ (\ref{eq:67})] alone. For numerically accessible versions of the integrals in
Eqs.\ (\ref{eq:73}, \ref{eq:75}) see Appendix B, Eqs.\ (\ref{eq:104}, \ref{eq:106}).

In the vicinity of the phase boundaries $I$ and $II$, one can evaluate the difference of $\Psi^{\rm ren}$
in the crystal and in the homogeneous phase analytically. The results are
\begin{equation}
\Psi^{\rm ren}_{\rm cryst} - \Psi^{\rm ren}_{\rm hom} \sim -C_I(T-T_0)^2 ,
\quad \Psi^{\rm ren}_{\rm cryst} - \Psi^{\rm ren}_{\rm hom} \sim C_{II}\frac{|T-T_0|}{\ln|T-T_0|},
\label{eq:76}
\end{equation}
respectively, where $T_0$ is the temperature at which the phase boundary is crossed.
In both cases, $C_I$ and $C_{II}$ are positive constants leading to $\Psi^{\rm ren}_{\rm cryst}<\Psi^{\rm ren}_{\rm hom}$.
The crystal is thermodynamically stable near the phase boundary, the phase transition is second order.

In order to prove Eq.\ (\ref{eq:76}) and to determine the constants one has to specify the direction in which one
enters the crystal phase. For simplicity we assume that we stay on a $\nu=$~const.\ line when passing through
the phase boundary. In this case we can use expansions (\ref{eq:72}) and (\ref{eq:74}) to calculate the
$\kk$-dependence of $a$, $s$ or $b$, and $\beta$ from Eqs.\ (\ref{eq:62}).
Although at $\kk=0$ one best works with $s$ instead of $b$ we use $s=1-\sn^{-2}(\ii b,\kk')$ and the last identity
of Eqs.\ (\ref{eq:74}) to give the result in terms of $b$. This unveils a remarkable similarity of the expansions at
$\kk=0$ and $\kk=1$. In either case we obtain
\begin{equation}
\hspace{-6mm}a=a_0+\frac{a_0}{4}z^2+a_2 z^4,\quad b=b_0+\frac{b_0}{4}z^2+b_2 z^4,
\quad\beta=\beta_0+\beta_2 z^4,\quad \hbox{with }z=\kk,\; \kk'.
\label{eq:77}
\end{equation}
By abuse of notation we used the same symbols for the expansion coefficients at $\kk=0$ and at $\kk=1$.
One has to remember that they assume different values depending on the phase boundary considered:
The constants $a_0$, $b_0$, $\beta_0$ obey Eqs.\ (\ref{eq:73}) or (\ref{eq:75}), respectively.
They are connected to the value of the effective fermion mass at the phase boundary by
$M^{(I)}_0=a_0\sqrt{s_0-1}/\beta_0$ or $M^{(II)}_0=a_0\sqrt{s_0}/\beta_0$.
The coefficients $a_2$ and $b_2$ become $\ln(\kk')$-dependent in the case $\kk\to1$.
In either limit they are too complex to write down. However, with the integrals
\begin{eqnarray}
J_0&=&\re\lim_{\e\to 0}\int_{-\infty+\ii\e}^{\infty+\ii\e}
 \dd \w \frac{\w^2}{\sqrt{\w^2-s_0+1}(\w^2-s_0)^3}\tanh\frac{a_0\w -\nu}{2}, \nonumber \\
J_1&=&\re\lim_{\e\to 0}\int_{-\infty+\ii\e}^{\infty+\ii\e}
 \dd \w \frac{\w^2}{(\w^2-s_0+1)^2\sqrt{\w^2-s_0}}\tanh\frac{a_0\w -\nu}{2} ,
\label{eq:79}
\end{eqnarray}
we can give the results for $\beta_2$,
\begin{eqnarray}
\beta_2^{(I)}&=&-\frac{a_0\beta_0J_0}{16(a_0s_0+\beta_0\gamma\sqrt{s_0-1})} ,\nonumber\\
\beta_2^{(II)}&=&-\frac{a_0\beta_0[1+4\ln(4/\kk')]J_1}{64(s_0-1)(a_0+b_0\beta_0\gamma)}.
\label{eq:77a}
\end{eqnarray}
Now we can use Eq.\ (\ref{eq:61}) or Eq.\ (\ref{eq:63}) to calculate $\Psi^{\rm ren}_{\rm cryst}$ as a function of $\kk$.
In order to compare the result with $\Psi^{\rm ren}_{\rm hom}$ in Eq.\ (\ref{eq:68}) [or  Eq.\ (\ref{eq:61}) with $\kk=0$ or $\kk=1$]
we have to determine the effective mass $M$ for given $\beta$, $\gamma$, $\nu$. Keeping in mind that Eq.\ (\ref{eq:67}) connects
$M$ to the above values of $a$, $s$, and $\beta$ at the phase boundary (and only at the phase boundary) this is achieved by using
$\partial_M \Psi^{\rm ren}_{\rm hom}=0$ [or by solving $F_0^{\rm ren}=F_1^{\rm ren}=0$ for $\kk=0$ or $\kk=1$ to (redundantly)
determine the values of $a$ and $s$ in the homogeneous phase and thereafter using Eq.\ (\ref{eq:67})].
We find that the coefficients given in Eq.\ (\ref{eq:77}) suffice to calculate $\Psi^{\rm ren}_{\rm cryst}-\Psi^{\rm ren}_{\rm hom}$
to the first non-vanishing order, yielding
\begin{eqnarray}
\pi\beta^2\left( \Psi^{\rm ren}_{\rm cryst}- \Psi^{\rm ren}_{\rm hom}\right)&=&\frac{a^2_0J_0}{512s_0}\kk^8+{\cal O}
(\kk^{10}),\nonumber\\
\pi\beta^2 \left(\Psi^{\rm ren}_{\rm cryst}- \Psi^{\rm ren}_{\rm hom}\right)&=&-\frac{a^2_0J_1}{16(s_0-1)}\kk'\,^4+{\cal O}
(\kk'\,^{6}).
\label{eq:80}
\end{eqnarray}
With $\kk^8=[(\beta-\beta_0)/\beta_2^{(I)}]^2$ and $\kk'\,^4=[(\beta-\beta_0)/\beta_2^{(II)}]$ we confirm
the behavior of Eq.\ (\ref{eq:76}). [Note that to leading order we have $1+4\ln(4/\kk')\sim\ln(\kk'\,^{-4})
\sim\ln[1/(\beta-\beta_0)]\sim\ln(T-T_0)$.] We see in Figs.\ 3, 4 that on a $\nu=$~const.\ line inside
the crystal phase the temperature is bounded by its values at the phase boundary:
$T_0^{(II)}<T<T_0^{(I)}$. Thus,
\begin{equation}
C_I=-\frac{a^2_0J_0}{512\pi s_0\beta^2_2T_0^2},\quad
C_{II}=\frac{4a_0(a_0T_0+b_0\gamma)}{\pi}\quad(\hbox{for }\gamma>0),
\label{eq:81}
\end{equation}
establishing $C_{II}>0$. Moreover, we deduce $\beta_2^{(I)}>0$ and by Eq.\ (\ref{eq:77a})
we conclude that $J_0<0$ yielding $C_I>0$.

In the massless case $\gamma=0$ the functional form of Eq.\ (\ref{eq:76}) is unchanged (see \cite{16}).
In the limit $\gamma\to0$ we have $s_0\to1$ and we observe that $C_I$ converges to the result
given in \cite{16} whereas $C_{II}\to4a_0^2T_0/\pi$ which is twice the value for $\gamma=0$.
In fact, the chiral limit is singular: The $\kk'\,^4$-coefficients of $\beta$ and of $\Psi^{\rm ren}_{\rm cryst}
- \Psi^{\rm ren}_{\rm hom}$ diverge in the limit $\gamma\to0$ which reflects the fact that both expressions are of order
$\kk'\,^2$ at $\gamma=0$. With this singularity we cannot expect that $C_{II}$ behaves smoothly
(or even that the functional form of Eq.\ (\ref{eq:76}) prevails). However, the massless and the massive case are
still related: In the derivation of $\beta_2^{(II)}$ the order in $\kk'$ enters linearly giving rise to the factor of two
in the massive case.

\vskip 0.1cm
{\em 6) The cusp}
\vskip 0.1cm
The phase boundary sheets $I$ and $II$ meet in a line labelled P$_{\rm t}$C in Fig.\ 3.
For $\gamma>0$ this line can be determined by taking the limit $b\to 0$.
As discussed in Subsect.~2.4, in this limit the scalar potential becomes both homogeneous and $\kk$-independent.
For fixed $\nu=\nu_{\rm c}$, solving $F_2=0$ gives $a$ as a function of $b$. The effective fermion mass
$M$ assumes a $\kk$-independent value $M_{\rm c}$ [see Eq.\ (\ref{eq:67}) with $s=1/b^2+{\cal O}(1)$],
\begin{equation}
M_{\rm c}=m_{\rm c}T_{\rm c},\quad m_{\rm c}=\lim_{b\to0}\frac{a(b)}{b},
\label{eq:82}
\end{equation}
where we have introduced the shorthand notation $m_{\rm c}$ and the cusp temperature $T_{\rm c}=1/\beta_{\rm c}$.
Because $M_{\rm c}$ is finite, we see that $a(b)$ vanishes in the limit $b\to0$, cancelling the $1/b$ pole in $\tilde{S}$.
However, $a$ is not a good expansion parameter at the cusp. Its expansion contains negative powers of $b$
[see Eq.\ (\ref{eq:89})]
and fails to converge for $b\to0$. The situation is reversed at $\gamma=0$. Here $b=\K$ in the whole $(\mu,T)$-plane
and the
cusp will be described by $a=0$. We conclude that at the multicritical point $M_{\rm t}=M_{\rm c}|_{\gamma=0}=0$
connects to the
massless Fermi gas phase as expected from the phase diagram. We will analyze the multicritical point by an expansion
at $a=0$ in the
next subsection. For the rest of this subsection we assume $\gamma>0$.

By introducing $m_{\rm c}$ and rescaling the integration variable $\w\to\w/b$ we obtain
\begin{eqnarray}
&&\hspace{-0.8cm}\left.F_0^{\rm ren}\right|_{b=0}=-\beta_c\gamma+\frac{m_{\rm c}}{2}\re\lim_{\e\to0}
 \int_{-\infty+\ii\e}^{\infty+\ii\e}\!\dd\w\frac{1}{(\w^2-1)^{3/2}}\tanh\frac{m_{\rm c}\w -\nu_c}{2}=0,\\
 \label{eq:83}
&&\hspace{-0.8cm}\left.F_1^{\rm ren}\right|_{b=0}=\gamma-\ln\beta_c+\lim_{\lambda\to\infty}\Biggl[
\ln\lambda -\frac{1}{2}\re\lim_{\e\to0} \int_{-\frac{\lambda}{2m_{\rm c}}+\ii\e}^{\frac{\lambda}{2m_{\rm c}}+\ii\e}
\!\dd\w \frac{\w^2}{(\w^2-1)^{3/2}}\tanh\frac{m_{\rm c}\w -\nu_c}{2}\Biggr]=0,\nonumber\\
&&\hspace{-0.8cm}\lim_{b\to0}\frac{15F_2/b^4}{\kk^2(1\!+\!\kk^2)-2u(1\!-\!\kk^2\!+\!\kk^4)}=I_2=\frac{1}{2}
\re\lim_{\e\to0} \int_{-\infty+\ii\e}^{\infty+\ii\e}\!\dd\w\frac{\w^2}{(\w^2-1)^{5/2}}\tanh\frac{m_{\rm c}\w -\nu_c}{2}
=0.\nonumber
\end{eqnarray}
By comparison with the homogeneous grand potential of Subsect.\ 4.3 given by Eq.\ (\ref{eq:68}) one finds (compare
Eq.\ (\ref{eq:70}) for the first equality)
\begin{eqnarray}
\left.\partial_M\pi\Psi^{\rm ren}_{\rm hom}\right|_{M=M_{\rm c}}&=&\frac{\left.F_0^{\rm ren}\right|_{b=0}}
{\beta_{\rm c}}+M_{\rm c}\left.F_1^{\rm ren}\right|_{b=0},\nonumber\\
\left.\partial^2_M\pi\Psi^{\rm ren}_{\rm hom}\right|_{M=M_{\rm c}}&=&\left.F_1^{\rm ren}\right|_{b=0},\nonumber\\
\left.\partial^3_M\pi\Psi^{\rm ren}_{\rm hom}\right|_{M=M_{\rm c}}&=&-\frac{3}{M_{\rm c}}I_2.
\label{eq:84}
\end{eqnarray}
We see that the minimum equations $F_i=0$ are equivalent to the conditions
$\Psi'=\Psi''=\Psi'''=0$ of the translationally invariant calculation \cite{13}.
This proves that the curve $b=0$ in the new phase diagram coincides with the line of critical points of the old phase
diagram.
We will see next that the opening angle of the cusp is zero (hence the name),
just like at $\gamma=0$ \cite{16}. Lying inside the crystal phase,
the critical first order phase transition sheet of the old phase diagram shown in Fig.~2 has then to be tangential to both
sheets $I$ and $II$.

In order to study the shape of the cusp in detail we expand the minimum equations at $b=0$.
After introducing the physical parameter $m_{\rm c}$ and rescaling $\w\to\w/b$ this is straightforward. We found the
following behavior
\begin{eqnarray}
T-T_{\rm c}&=& C_1(\mu-\mu_c) + C_{3/2}(\kk)(\mu-\mu_{\rm c})^{3/2}+C_2(\kk)(\mu-\mu_{\rm c})^2 + \ldots \, ,
\label{eq:85} \\
C_1&=&\frac{I_{\rm c} T_{\rm c}}{I_{\rm c} \mu_{\rm c}+\gamma/m_c-T_{\rm c}} , \quad
I_{\rm c} = \frac{1}{4}\re\lim_{\e\to 0}\int_{-\infty +\ii \e }^{\infty+\ii \e}
\frac{\dd\w}{\sqrt{\w^2-1}}\cosh^{-2}\frac{m_{\rm c}\w -\nu_c}{2}.\nonumber
\end{eqnarray}
The linear coefficient $C_1$ is independent of $\kk$ which proves that the phase boundaries $I$ with $\kk=0$ and
$II$ with $\kk=1$ are parallel at the cusp.
 
For $\gamma\to 0$ we have $b\to\K$ in the $(\mu,T)$-plane. Hence small values of $b$ are confined to smaller and
smaller regions about the cusp when $\gamma$ goes to zero.
We therefore cannot expect that the expansion (\ref{eq:85}) valid for positive $\gamma$ converges to the expansion for
$\gamma=0$. Comparison with \cite{16} actually shows that $C_1$ converges to the chiral value which means that the
orientation of the cusp changes continuously with $\gamma\to0$. Moreover, $C_{3/2}$ vanishes in the chiral limit which
agrees with the absence of this coefficient for $\gamma=0$. However, $C_2$ fails to converge to the chiral result.
 
The $\gamma$-dependence of the cusp can most clearly be seen in the two-dimensional sections of Fig.~4.

\vskip 0.1cm
{\em 7) Multicritical point and Ginzburg-Landau effective action}
\vskip 0.1cm
The most remarkable of the special points in Table 1, Sect.~4.4, is the multicritical
point P$_{\rm t}$ at ($\gamma=0,\mu_{\rm t}=0.608221,T_{\rm t}=0.318329$).
At this point the value of $a$ goes to zero and we can expand all relevant functions in $a$. The expansion coefficients 
are given
in terms of the polygamma function. This allows us to give an approximate solution of the minimum equations $F_i=0$ 
valid in the neighborhood
of the critical point for small values of $\gamma$. Moreover, we will derive a Ginzburg-Landau effective action at the 
end of this
subsection.

The method to obtain the series in $a$ is identical to the chiral case in \cite{16}.
After a rescaling $\w\to\w/a$ the expansion in $a$ is obtained from a large $\w$ series of the algebraic part of the integrand.
We define the coefficients $c_n$ by a generating function,
\begin{equation}
\frac{\w^3}{\sqrt{W}} =  \sum_{n=0}^\infty \frac{c_n}{\w^{2n}} .
\label{eq:86}
\end{equation}
We find $c_0=1$, $c_1=(3s-1-\kk^2)/2$. Note that the $c_n$ fulfill the recursion relation
\begin{eqnarray}
2(n+1)c_{n+1}&=&(2n+1)(3s-1-\kk^2)c_n-2n(3s^2-2s(1+\kk^2)+\kk^2)c_{n-1}\nonumber\\
&&+\;(2n-1)(s-1)(s-\kk^2)s\,c_{n-2}.
\label{eq:87}
\end{eqnarray}
Now, we mainly use the residue theorem to obtain the following identities valid in the limit $\lambda\to\infty$
\begin{equation}
a\re\lim_{\e\to0}
 \int\limits_{-\frac{\lambda}{2a}+\ii\e}^{\frac{\lambda}{2a}+\ii\e}\frac{\dd\w}{\w^{2n}}\ln\left(2\cosh
\frac{a\w-\nu}{2}\right)
=\left\{
\begin{array}{ll}
\displaystyle\frac{\lambda^2}{8}+\frac{\nu^2}{2}+\frac{\pi^2}{6}&n=0\\
\displaystyle a^2\left[\ln\frac{\lambda}{4\pi}-1-\re\,\psi\left(\frac{1}{2}+\frac{\ii\nu}{2\pi}\right)\right]&n=1\\
\multicolumn{2}{l}{
\displaystyle\frac{4\pi^2}{(2n-1)!}\left(\frac{-a^2}{4\pi^2}\right)^n\re\,\psi\left(2n-2,\frac{1}{2}+\frac{\ii\nu}{2\pi}\right)}\\
&n\geq2.
\end{array}\right.
\label{eq:88}
\end{equation}
Expansions of integrands with factor $\tanh[(a\w-\nu)/2]$ are obtained by differentiation with respect to $a$ or $\nu$.
From Eqs.\ (\ref{eq:61}, \ref{eq:62}) we derive the following expressions $(c_{-1}=0)$
\begin{eqnarray}
F_0^{\rm ren}&=&-\beta\gamma-at\sum_{n=1}^\infty\frac{c_{n-1}}{(2n)!}\left(\frac{-a^2}{4\pi^2}\right)^n
 \re\,\psi\left(2n,\frac{1}{2}+\frac{\ii\nu}{2\pi}\right),\nonumber\\
F_1^{\rm ren}&=&\gamma-\ln\frac{\beta}{4\pi}+\sum_{n=0}^\infty\frac{c_n}{(2n)!}\left(\frac{-a^2}{4\pi^2}\right)^n
 \re\,\psi\left(2n,\frac{1}{2}+\frac{\ii\nu}{2\pi}\right),\nonumber\\
F_2&=&\sum_{n=1}^\infty
 \frac{-2n(s-u)c_n+a_1c_{n-1}+a_2c_{n-2}}{(2n+1)!}\left(\frac{-a^2}{4\pi^2}\right)^n
 \re\,\psi\left(2n,\frac{1}{2}+\frac{\ii\nu}{2\pi}\right),\nonumber\\
 && a_1=-(2n+1)t\Z+2ns(2s-1-\kk^2)-(s-1)(s-\kk^2),\nonumber \\
 && a_2=-(2n-1)(s-1)(s-\kk^2)s,\nonumber\\
 \label{eq:89}
\pi\beta^2\Psi^{\rm ren}&=&-\frac{\nu^2}{2}-\frac{\pi^2}{6}-\beta^2\gamma\langle S\rangle\\
&&+\;a^2\sum_{n=0}^\infty
 \frac{c_{n+1}-(s-u)c_n}{(2n+1)!}\left(\frac{-a^2}{4\pi^2}\right)^n\!\left[
 \re\,\psi\left(2n,\frac{1}{2}+\frac{\ii\nu}{2\pi}\right)+\delta_{n,0}\left(\gamma-\ln\frac{\beta}{4\pi}\right)\right]\!.\nonumber
\end{eqnarray}
These expansions can be used to solve the minimum equations to lowest order in $a$. The result is (one may
conveniently use Eq.\ (\ref{eq:63}) for the last identity)
\begin{eqnarray}
\frac{a^2}{4\pi^2}&\sim&A(\kk)\frac{\re\,\psi\left(2,\frac{1}{2}+\frac{\ii\nu}{2\pi}\right)}
 {\re\,\psi\left(4,\frac{1}{2}+\frac{\ii\nu}{2\pi}\right)},\nonumber\\
A(\kk)&=&\frac{40[3t\Z+\kk^2-(3s-1-\kk^2)u]}
 {5t\Z[3s\!-\!1\!-\!\kk^2]+\kk^2[5s\!-\!1\!-\!\kk^2]-[15s^2\!-\!10s(1\!+\!\kk^2)\!+\!3\!+\!2\kk^2\!+\!3\kk^4]u},\nonumber\\
\ln\frac{\beta}{4\pi}&\sim&\re\,\psi\left(\frac{1}{2}+\frac{\ii\nu}{2\pi}\right)-\frac{A(\kk)}{2}\left(c_1-c_2\frac{A(\kk)}
{12}\right)
 \frac{\left[\re\,\psi\left(2,\frac{1}{2}+\frac{\ii\nu}{2\pi}\right)\right]^2}
 {\re\,\psi\left(4,\frac{1}{2}+\frac{\ii\nu}{2\pi}\right)},\nonumber\\
\gamma&\sim&\frac{\pi tA(\kk)^{3/2}}{\beta}\left(1-c_1\frac{A(\kk)}{12}\right)
 \frac{\left[\re\,\psi\left(2,\frac{1}{2}+\frac{\ii\nu}{2\pi}\right)\right]^{5/2}}
 {\left[\re\,\psi\left(4,\frac{1}{2}+\frac{\ii\nu}{2\pi}\right)\right]^{3/2}},
\label{eq:90}\\
\pi\beta^2\Psi^{\rm ren}_\mathrm{min}&\sim&-\frac{\nu^2}{2}-\frac{\pi^2}{6}-2\pi^2
A(\kk)^2 \left( c_1^2-c_2+(c_3-c_1c_2)\frac{A(\kk)}{12}\right)
\frac{\left[\re\,\psi\left(2,\frac{1}{2}+\frac{\ii\nu}{2\pi}\right)\right]^3}
{\left[\re\,\psi\left(4,\frac{1}{2}+\frac{\ii\nu}{2\pi}\right)\right]^2}.
\nonumber
\end{eqnarray}
The approximation is improved if one corrects $T$ to $T\approx\exp(-\gamma)/\beta$ with the above values for $\beta$ and $\gamma$.
If $\nu$ is close to $\nu_{\rm t}$ then $\re\,\psi(2,\frac{1}{2}+\frac{\ii\nu}{2\pi})$ is small and
for $\nu>\nu_{\rm t}$ the above equations give a ($b,\kk,\nu$)-parameter
description of the crystal phase for low $\gamma$ near the multicritical point. It gives insight into the behavior of the critical line
P$_{\rm t}$C for small $\gamma$. We read off $\gamma={\cal O}(\Delta T^{5/4})={\cal O}(\Delta\mu^{5/4})$. A systematic extension
to higher orders of $\re\,\psi(2,\frac{1}{2}+\frac{\ii\nu}{2\pi})$ is possible.

We close this subsection with the observation that in the vicinity of the Lifshitz point, we can derive a Ginzburg-Landau
effective action from Eq.\ (\ref{eq:89}). A similar calculation was done at $\gamma=0$ and
found to agree with independent calculations in condensed matter physics \cite{4,35}. Here we are
interested in a generalization to small but finite values of $\gamma$.
Using Eq.\ (\ref{eq:20}) one can verify that the coefficients $c_{n+1}-(s-u)c_n$ in Eq.~(\ref{eq:89}) are related to the following
combinations of spatial averages of powers of $S$, Eqs.~(\ref{eq:5}), and its derivatives,
\begin{eqnarray}
c_1-(s-u) c_0 & = & \frac{1}{2}\langle\tilde{U}_-\rangle\;=\; \frac{1}{2A^2} \langle S^2 \rangle  \nonumber \\
\label{eq:91}
c_2-(s-u) c_1 & = &\frac{3}{8}\langle\tilde{U}_-^2\rangle\;=\; \frac{3}{8A^4}\left(\langle S^4 \rangle +\langle (S')^2\rangle \right) \\
c_3-(s-u) c_2 & = & \frac{5}{16}\left(\langle\tilde{U}_-^3\rangle+\frac{1}{2}\langle\tilde{U}_-^{\prime\,2}\rangle\right)
\;=\;\frac{5}{16A^6} \left(\langle S^6\rangle+5\langle S^2(S')^2 \rangle+\frac{1}{2}\langle(S'')^2\rangle \right) .\nonumber
\end{eqnarray}
Although the $c_n$ and $S(x)$ are different from the corresponding quantities in the chiral limit, the fact that
Eqs.~(\ref{eq:91}) hold enables us to write down the Ginzburg-Landau effective action in the simple form
\begin{equation}
\Psi_{\rm eff} =\left.\Psi_{\rm eff}\right|_{\gamma=0}+\frac{\gamma}{2\pi}\left(S^2 - 2S\right),
\label{eq:92}
\end{equation}
where the effective action at $\gamma=0$ is the same as in Ref.~\cite{16},
\begin{eqnarray}
  \Psi_{\rm eff}(\gamma=0) & =&  - \frac{\pi}{6} T^2 - \frac{\mu^2}{2\pi} + \frac{1}{2\pi}  S^2  \left[ \ln (4\pi T)+
{\rm Re}\  \psi \left( \frac{1}{2} +
 \frac{\ii\mu}{2\pi T}\right) \right] \nonumber \\
  & -& \frac{1}{2^6 \pi^3 T^2} \left(  S^4  +  S'\,^2  \right) \re \  \psi \left( 2, \frac{1}{2} + \frac{\ii\mu}{2\pi T}\right) \nonumber \\
&   & + \frac{1}{2^{11}3\pi^5 T^4} \left(  S^6  + 5  S^2 S'\,^2  + \frac{1}{2}  S''\,^2 \right) \re\  \psi \left( 4, \frac{1}{2} +
 \frac{\ii\mu}{2\pi T}\right) .
\label{eq:93}
\end{eqnarray}
This action can systematically be extended to higher orders in $S$.
The instability with respect to crystallization is related to the fact that the ``kinetic" $(S^4+S'\,^2)$-term
can change sign, depending on $\mu$ and $T$.

\vskip 0.1cm
{\em 8) Zero temperature limit}
\vskip 0.1cm
The massive GN model at zero temperature was solved in \cite{12}. Here we give a short account
of how to regain the results in our general formalism.

In the limit $T\to0$ both $a$ and $\nu$ go to infinity. The ratio $\nu/a=\mu/A$, however, tends towards a well defined
limit which lies in the gap,
\begin{equation}
\sqrt{s-\kk^2}<\lim_{T\to0}\frac{\nu}{a}<\sqrt{s}.
\end{equation}
Moreover, we observe that $\ln 2\cosh[(a\w-\nu)/2]\to|(a\w-\nu)/2|$ and $\tanh[(a\w-\nu)/2]\to\hbox{sgn}(a\w-\nu)$.
From an analytic point of view, this splits the integration contour into two sections $\re\,\w>\nu/a$ and $\re\,\w<\nu/a$.
The orientation of the second section may be reversed to compensate for the relative minus sign and
then deformed to match the first. In addition we obtain a big half circle to account for the singularities at infinity.
Considering that we are interested in the real part only we may
move the left end of the contour from $\nu/a$ to $\sqrt{s}$, the right end of the gap.
This (plus a partial integration for $F_2$) transforms the integral representations of $\Psi$ and the $F_i$ to
\begin{eqnarray}
\left.\pi\Psi^{\rm ren}\right|_{T=0}&\!\!=\!\!&\frac{\Lambda^2}{8}-\frac{\mu\Lambda}{2} +\frac{A^2 \langle
\tilde{S}^2\rangle}{2} \hspace{-1pt}(\ln\Lambda\!+\!\gamma\!-\!1) - A\gamma \langle\tilde{S}\rangle
-\!\int_{\sqrt{s}}^\frac{\Lambda}{2A} \hspace{-3pt}\dd \w\frac{A\w(\w^2\!-\!s\!+\!u)(A\w\!-\!\mu)}{\sqrt{W}} \hspace{-1pt},\nonumber\\
\lim_{T\to0}TF_0^{\rm ren}&\!\!=\!\!&-\gamma+At\int_{\sqrt{s}}^\infty \dd \w \frac{1}{\sqrt{W}},\nonumber\\
\left.F_1^{\rm ren}\right|_{T=0}&\!\!=\!\!&\gamma+\ln \Lambda
-\int_{\sqrt{s}}^\frac{\Lambda}{2A} \dd \w \frac{\w^2}{\sqrt{W}},\\
\left.F_2\right|_{T=0}&\!\!=\!\!&-\Lambda^2+\frac{\mu\Lambda}{A}+\frac{\langle\tilde{S}^2\rangle}{2}
+\int_{\sqrt{s}}^\Lambda \dd \w  \Biggl[\frac{\w (\w^2\!-\!s\!+\!u)(A\w-\mu)} {A\sqrt{W}}
+\frac{t \Z\! -\!u\w^2}{\sqrt{W}} + \frac{\sqrt{W}}{\w^2\!-\!s}\Biggr],\nonumber
\end{eqnarray}
with the limit $\Lambda\to\infty$ understood.
After the substitution $\w^2=z$ these integrals evaluate in a standard way to incomplete elliptic integrals.
The result can be extracted from \cite{12}.
\section{Summary and outlook}
The motivation for generalizing the GN model to finite bare fermion masses is twofold. On the one hand,
from the point of view of strong interaction physics, quarks are massive and chiral symmetry is
explicitly broken in QCD. On the other hand, a prominent application of the GN model
in condensed matter physics are conjugated conducting polymers like doped polyacetylene. Here, the difference
between massless and massive GN models (with unbroken or broken discrete chiral symmetry) translates
into polymers with degenerate or non-degenerate ground states, e.g., {\em trans}- and {\em cis}-polyacetylene.
Since non-degenerate polymers prevail in nature, the model with broken symmetry 
is of considerable interest as evidenced by numerous theoretical works on the bipolaron lattice in the 80's and 90's.

In this paper we have discussed the thermodynamics of the massive GN model in the large $N$ limit.
Our main result is the exact, comprehensive phase diagram of this model as a function of temperature and
chemical potential for arbitrary
bare fermion mass. Previous attempts to solve the same problem have only considered translational invariant
condensates and produced results which fail theoretical tests.
They miss in particular the Peierls effect which provides a simple physical explanation for the emergence
of a crystalline state in one-dimensional fermion systems, namely gap formation at the Fermi surface.

The phase diagram in Fig.~3 brings a series of recent investigations
to a close: The $T=0$ coordinate plane
was the subject of Ref.~\cite{12} where we addressed the issue of cold, dense matter in the massive
GN model. The $\gamma=0$ coordinate plane, i.e., the chiral limit, has been investigated first numerically \cite{14}
and later analytically \cite{16}. The intersection of these two planes, the $\mu$-axis, was the
subject of Ref.~\cite{15}.
In these various limits integrals which had to be evaluated numerically
in the present work simplify. At $T=0$ in particular, they can be reduced to complete (at $\gamma=0$) or incomplete
(at $\gamma>0$) elliptic integrals. The results in the various limits remain of interest since they give additional
analytical insights.

As mentioned before, the GN model is a good example for a fruitful interaction between particle and
condensed matter physics. At first, semi-classical methods developed for baryons in the relativistic case
have been taken over to describe solitons and polarons in polymer physics. In the following years, condensed matter
theory made significant progress in the treatment of periodic arrays of polarons. This aspect was
overlooked in relativistic field theory for a long time, presumably because crystals are not
common in particle physics. The full phase diagram of the massive model
presented here may well have applications to real systems in condensed matter physics again. Particle physics
and condensed matter physics get integrated once one addresses problems of fermionic matter and
thermodynamics. Incidentally, this trend can also be observed in QCD where interest in
color superconductivity has led to a similar exchange of ideas \cite{36}.

\section*{Appendix A: Coordinate transformation}
The Jacobian matrix for the change of coordinates $b,A,\kk \to  \langle S \rangle, \langle S^2 \rangle, \ell$
in Eq.~(\ref{eq:57}) has the following entries,
\begin{eqnarray}
\frac{\partial \langle S \rangle}{\partial b} & = & -A(s-u),
\nonumber \\
\frac{\partial \langle S^2 \rangle}{\partial b} & = & -2tA^2,
\nonumber \\
\frac{\partial \langle \ell \rangle}{\partial b} & = & 0,
\nonumber \\
\frac{\partial \langle S \rangle}{\partial A} & = & \Z+\frac{t}{s},
\nonumber \\
\frac{\partial \langle S^2 \rangle}{\partial A} & = & 2A(s-1-\kk^2+2u),
\nonumber \\
\frac{\partial \langle \ell \rangle}{\partial A} & = & -\frac{2{\K}}{A^2},
\nonumber \\
\frac{\partial \langle S \rangle}{\partial \kk} & = & \frac{A\left[\Z(s-\kk^2)-b(u-\kk^2)(s-u)-\kk^2 t/s \right]}
{\kk (1-\kk^2)},
\nonumber \\
\frac{\partial \langle S^2 \rangle}{\partial \kk} & = & \frac{2A^2\left[t\Z-\kk^2(s-1)+(u-\kk^2)(u-\kk^2-bt)\right]}
{\kk (1-\kk^2)},
\nonumber \\
\frac{\partial \langle \ell \rangle}{\partial \kk} & = & -\frac{2(u-\kk^2){\K}}{A\kk(1-\kk^2)}.
\label{eq:94}
\end{eqnarray}
The parameters $s$, $t$, $u$ are defined in Eq.~(\ref{eq:24}) and $\Z=\Z(b,\kk)$ is Jacobi's Zeta function.
The inverse Jacobian matrix will be denoted by $M$ and its entries are given by
\begin{eqnarray}
M_{11} & = & \frac{t\Z-u(s+u-1-\kk^2)-bt(u-\kk^2)}{Au(1-u)(u-\kk^2)},
\nonumber \\
M_{12} & = & -\frac{t}{u(1-u)},
\nonumber \\
M_{13} & = & \frac{\kk (1-\kk^2)t}{Au(1-u)(u-\kk^2)},
\nonumber \\
M_{21} & = & \frac{s(s-u)[-\Z+b(u-\kk^2)]+tu}{2A^2su(1-u)(u-\kk^2)},
\nonumber \\
M_{22} & = & \frac{s-u}{2Au(1-u)},
\nonumber \\
M_{23} & = & -\frac{\kk (1-\kk^2)(s-u)}{2A^2u(1-u)(u-\kk^2)},
\nonumber \\
M_{31} & = & \frac{\Z [t\Z\!-\!2su\!+\!u^2\!+\!\kk^2\!-\!bt(u\!-\!\kk^2)]
+b(u\!-\!\kk^2)[u(s\!+\!1\!+\!\kk^2)\!-\!2u^2\!-\!\kk^2]+u^2t/s}{u(1-u)(u-\kk^2)\ell},
\nonumber \\
M_{32} & = & -\frac{A[t\Z-(s+\kk^2)u+u^2+\kk^2]}{u(1-u)\ell},
\nonumber \\
M_{33} & = & \frac{\kk(1-\kk^2)[t\Z-u(s+1+\kk^2)+2u^2+\kk^2]}{u(1-u)(u-\kk^2)\ell}.
\label{eq:95}
\end{eqnarray}
In the derivation of Eqs.~(\ref{eq:95}) one has to use the relation between $t^2$ and $s$ given in Eq.~(\ref{eq:25}).

On the left hand side of Eq.\ (\ref{eq:57}) we need the derivatives of $\pi \beta^2 \Psi_1$, Eq.~(\ref{eq:50}),
with respect to the original variables $b$, $A$, $\kk$. One finds
\begin{eqnarray}
\frac{\partial}{\partial b} \pi \beta^2 \Psi_1 &=& a^2 t  -at \,{\rm Re} \lim_{\e \to 0}\int_{\Lambda_{\w}+
{\rm i}\e}^{\infty +{\rm i}\e}{\rm d}\w \left( \frac{\partial}{\partial \w}
\frac{(\w^2-s+u)}{\sqrt{W}}\right) \ln \left( 1+ {\rm e}^{-a\w + \nu}\right),
\nonumber \\
\frac{\partial}{\partial A}\pi \beta^2 \Psi_1 & = & - \frac{\beta^2}{A}  \langle S^2 \rangle
+ \beta \,{\rm Re} \lim_{\e \to 0}\int_{\Lambda_{\w}+
{\rm i}\e}^{\infty +{\rm i}\e}{\rm d}\w \left( \frac{\partial}{\partial \w}
\frac{\w(\w^2-s+u)}{\sqrt{W}}\right)\w \ln \left( 1+ {\rm e}^{-a\w + \nu}\right),
\nonumber \\
\frac{\partial}{\partial \kk} \pi \beta^2 \Psi_1 & = & - \frac{\beta^2}{2} \frac{\partial \langle S^2 \rangle}
{\partial \kk}- a \,{\rm Re} \lim_{\e \to 0}\int_{\Lambda_{\w}+
{\rm i}\e}^{\infty +{\rm i}\e}{\rm d}\w \left[ - \frac{1}{2}\frac{\partial s}{\partial \kk}
\left( \frac{\partial}{\partial \w} \frac{(\w^2-s+u)}{\sqrt{W}} \right) \right.
\nonumber \\
& &\left.+\; \frac{\partial u}{\partial \kk} \frac{\w}{\sqrt{W}}- \frac{\kk \w(\w^2-s+u)}
{(\w^2-s+\kk^2)\sqrt{W}}\right] \ln \left( 1+ {\rm e}^{-a\w + \nu}\right),\nonumber\\
\hbox{with}&&\frac{\partial u}{\partial \kk}=\frac{(u-\kk^2)^2+\kk^2(1-\kk^2)}{\kk(1-\kk^2)},\quad
\frac{\partial s}{\partial \kk}=\frac{1}{A^2}\frac{\partial \langle S^2\rangle}{\partial \kk}
+2\kk-2\frac{\partial u}{\partial \kk}.
\label{eq:96}
\end{eqnarray}
The terms which do not involve an integral arise from differentiating the cutoff
$\Lambda_{\w}$. In the other terms, we differentiate the integrand. When taking the
derivative with respect to $A$, we first change integration variables $\w \to \w/a$ to
avoid having to differentiate the $\ln$-factor. Apart from that, we have made repeated use of
the structure of the prefactor of $\ln\left(1+{\rm e}^{-a\w + \nu}\right)$ in the integrand
to replace various derivatives by derivatives with respect to $\w$.

We can now invert Eq.~(\ref{eq:57}) by applying the matrix $M$ of Eq.~(\ref{eq:95}) to the vector (\ref{eq:96})
and adding the contribution from $\Psi_2$, Eq.\ (\ref{eq:56}). The result can be cast into the compact form
given in Eq.~(\ref{eq:58}) as can easily be verified with the help of computer algebra.

\section*{Appendix B: Formulae used in the numerical calculations}

In this appendix we transform contour integrals into real integrals of smooth integrands with compact support.
This step makes the integrals accessible to numerical evaluation.

The contour of integration of the original integrals is shifted slightly above the real axis, with a symmetric cutoff $\Lambda'$
at $\pm\infty$ if needed. The integration variable is $\w$.
The integrand $f$ has cuts or singularities on the real line, but no pole at the origin
(see Fig.\ 1 in Subsect.\ 3.2). As a function of $\nu$ it has the
symmetry $f(-\w,\nu)=\pm f(\w,-\nu)$, depending on the location of the cuts.
We are only interested in the real part of the integral.
The transformation of the integrals follows the steps
(see Fig.\ 6; some steps may be skipped, depending on the integral considered)

\vskip 0.2cm
\begin{figure}[ht]
\begin{center}
\epsfig{file=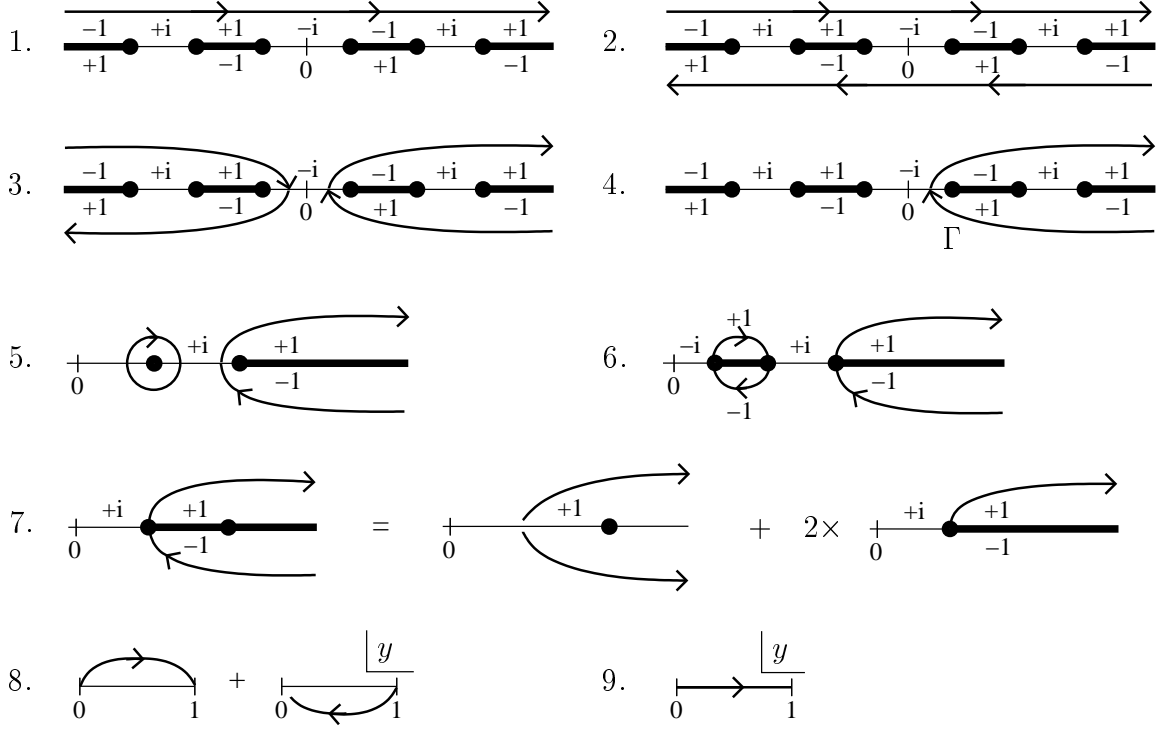}
\vskip 0.2cm
\caption{Successive conversion of contour integrals into real integrals with compact support.
The numbering refers to the corresponding steps explained in the main text.} 
\end{center}
\end{figure}
\begin{enumerate}
\item
Swap gaps with bands: We may swap the cuts with the intervals not cut on the real axis without
affecting the function in the upper half plane. We thus avoid a branch cut at the origin.
\item
Symmetrize the contour: We split the integral into two halves and move the contour of one half below the real axis.
Due to the branch cuts, we acquire a minus sign which
we use to invert the orientation of the contour. As a byproduct the imaginary parts of the integrals cancel,
$\re\int_{-\Lambda'+\ii\e}^{\Lambda'+\ii\e}\dd\w f
=\frac{1}{2}(\int_{-\Lambda'+\ii\e}^{\Lambda'+\ii\e}+\int_{\Lambda'-\ii\e}^{-\Lambda'-\ii\e})\dd\w f$.
\item
Cut and paste the contour at the origin:\\
$\int_{-\Lambda'+\ii\e}^{\Lambda'+\ii\e}+\int_{\Lambda'-\ii\e}^{-\Lambda'-\ii\e}
=(\int_{-\Lambda'+\ii\e}^0+\int_0^{-\Lambda'-\ii\e})+(\int_{\Lambda'-\ii\e}^0+\int_0^{\Lambda'+\ii\e})$.
\item
Reflect the left path: By symmetry of the contours and a sign change in the measure $\dd\w$ we are left with
integrating $f(\w,\nu)-f(-\w,\nu)$ along the right path $\Gamma=\Lambda'-\ii\e\to0\to\Lambda'+\ii\e$.
Using $f(-\w,\nu)=\pm f(\w,-\nu)$ we have arrived at
\begin{equation}
\re\,\int_{-\Lambda'+\ii\e}^{\Lambda'+\ii\e} f(\w,\nu)=\frac{1}{2}\int_\Gamma f(\w,\nu)\mp(\nu\to-\nu),\quad
\quad f\hbox{ holomorphic on }\Gamma.
\label{eq:97}
\end{equation}
\item
Isolated poles: Isolated poles are evaluated with the residue theorem.
\item
Shrink the contour: The contour is shrunk to embrace the cuts.
\item
Separate pole contributions: Poles on cuts are subtracted (here, in a symmetric way with respect to $\w\to-\w$)
and evaluated separately by elementary integration. Note that by the sign change at the cut the subtraction coefficients
have opposite signs for the arcs above and below the real axis. This is indicated by reversing the orientation of the
lower arc in Fig.\ 6. Now that the cuts are free of additional singularities the integrand assumes opposite signs above
and below the real axis. The value of the integral is twice the integral over the upper arc stretching from left to right end
of the cut, or ending at $\Lambda'+\ii\e$.
\item
Parametrize the cuts: We are left with one or two integrals over (possibly half open)
cuts. These intervals are parameterized algebraically. For a cut $\w\in \w_1\dots \w_2$ we use
\begin{equation}
\w^2=\w_1^2+(\w_2^2-\w_1^2)\Delta^2\quad\hbox{with}\quad\Delta=\frac{2y}{1+y^2}\quad\hbox{and}\quad y=0\dots1.
\label{eq:98}
\end{equation}
For a half open cut $\w\in \w_3\dots\Lambda'+\ii\e$ we take
\begin{equation}
\w^2=\w_1^2+(\w_3^2-\w_1^2)\Delta^{-2}\quad\hbox{with}\quad y=1\dots
\frac{\sqrt{(\Lambda'+\ii\e)^2-\w_1^2}-\sqrt{(\Lambda'+\ii\e)^2-\w_3^2}}{\sqrt{\w_3^2-\w_1^2}}
\label{eq:98a}
\end{equation}
and $\w_1=0$ in case of a single cut.
This choice of parameterization lifts the square root singularities at the ends of the cut and has the additional advantage
to produce squares in $W$, thus simplifying $\sqrt{W}$.
\item
Regularize the limit $\Lambda'\to\infty$: If necessary we subtract powers of $1/y$ and evaluate these terms in the limit
$\Lambda'\to\infty$ by elementary integration. This renders the integral finite at $y=0$. After swapping the orientation
we straighten the contour and combine the integrand with a possible second contribution and all constants to obtain a
single real integral ranging from zero to one.
\item
(optional) Modification for improved numerics: At some instances we have to find zeros of functions containing
integrals. In this case it is useful to present the function as integral over an
integrand vanishing at the ends of the integration interval. To this end we add the polynomial
$-f(0)(3y^2-4y+1)-f(1)(3y^2-2y)$ to the integrand $f(y)$.
Moreover, the integrands may have a lifted singularity which may or may not need extra attention depending on the method of
numerical integration. In any case it is possible to analytically determine the position of the lifted singularity and the (finite)
value of the integrand at that point.
\end{enumerate}

In the following we list the integrands $f(y)$, $y=0\ldots1$, following the above steps 1 to 9 (without step 10)
for various integrals of 1) the crystal phase, 2) the phase boundary $I$, and 3) the phase boundary $II$.

\vskip 0.1cm
{\em 1) Eqs.\ (\ref{eq:61}, \ref{eq:62}): The crystal phase}
\begin{eqnarray}\label{eq:99}
\hbox{With}&&f_n=2\frac{T_2^{n-1}\ln\left(2\cosh\frac{aT_2-\nu}{2}\right)-
T_1^{n-1}\ln\left(2\cosh\frac{aT_1-\nu}{2}\right)}{(1+y^2)\sqrt{1-(1-\kk^2)\Delta^2}} -(-1)^n(\nu\to-\nu),\nonumber\\
&&\nonumber\\
&&T_1^2=s-1+(1-\kk^2)\Delta^2,\quad T_2^2=s-1+\Delta^{-2},\quad\hbox{we obtain}\nonumber\\
&&\nonumber\\
\pi\beta^2\Psi^{\rm ren}\!\!&:&-a[f_3-(s\!-\!u)f_1+\beta\gamma\langle\tilde{S}\rangle]
 +\frac{a^2}{8}\left[\frac{2}{y^3}+4s-1\right]+\frac{a^2\langle\tilde{S}^2\rangle}{2}
 \left[\frac{1}{y}+\ln\frac{a}{\beta}+\gamma-1\right],\nonumber\\
I_0\!\!&:&-\partial_\nu f_0,\nonumber\\
I_1\!\!&:&-\partial_\nu f_2-\frac{1}{y}-\ln a,\nonumber\\
F_2\!\!&:&\frac{f_3}{a}-(s\!-\!u)\left[\frac{f_1}{a}-\partial_\nu f_2\right]-\partial_\nu f_4
+\langle \tilde{S}^2\rangle\left[\partial_\nu f_2+\frac{1}{2}\right]
-\left[t\Z+(s\!-\!1)(s\!-\!\kk^2)\right]\partial_\nu f_0\nonumber \\
&&-\frac{1}{4}\left[\frac{2}{y^3}+4s-1\right].
\end{eqnarray}

\vskip 0.1cm
{\em 2) Eqs.\ (\ref{eq:73}, \ref{eq:79}): Phase boundary I}
\begin{eqnarray}\label{eq:104}
\hbox{With}&&g_n=\frac{{\rm s}\,{\rm c}^n\Delta^{-n}y\tanh\left(\frac{\alpha}{2\Delta}-\frac{\nu}{2}\right)
 +{\rm c}(\Delta^{-1}y-1)\tanh\left(\frac{\alpha}{2{\rm c}}-\frac{\nu}{2}\right)}{2y^2{\rm s}^{n-1}({\rm c}^2\Delta^{-2}-1)}\nonumber\\
&&\quad\quad\;+\frac{1}{2{\rm s}^{n-1}}\ln\frac{1+\rm c}{\rm s}
\tanh\left(\frac{\alpha}{2{\rm c}}-\frac{\nu}{2}\right)+(-1)^n(\nu\to-\nu),\nonumber\\
&&\nonumber\\
&&{\rm c}=\cos b, \qquad {\rm s} = \sin b, \qquad\hbox{we obtain}\nonumber\\
&&\nonumber\\
\left.I_0\right|_{\kk=0}\!\!&:&\left.g_0\right|_{\alpha=a\,{\rm c/s}},\nonumber\\
\left.I_1\right|_{\kk=0}\!\!&:&\left.g_2-\frac{\Delta}{y^2}+\frac{1}{y}-\ln2\alpha\right|_{\alpha=a\,{\rm c/s}},\nonumber\\
\lim_{\kk\to0}\frac{16s}{\kk^4}F_2\!\!&:&\left.{\rm s\,c}\,\partial_b\,g_2\right|_{\alpha=a\,{\rm c/s}},\nonumber\\
J_0\!\!&:&\left.\frac{{\rm s}^3}{4\rm c}\,\partial_b\frac{{\rm c}^3}{\rm s}\,\partial_b\,g_2\right|_{\alpha=a_0\,{\rm c/s},\,b=b_0}.
\end{eqnarray}

\vskip 0.1cm
{\em 3) Eqs.\ (\ref{eq:75}, \ref{eq:79}): Phase boundary II}
\begin{eqnarray}\label{eq:106}
\hbox{With}&&h_n=\frac{{\rm sh}^{2-n}\,{\rm ch}^n\Delta^{-n}}{y({\rm ch}^2\Delta^{-2}-1)}\ln\left[2\cosh\left(
\frac{\alpha}{2\Delta}-\frac{\nu}{2}\right)\right]\nonumber\\
&&\quad\quad\;-\frac{\pi}{2}{\rm sh}^{1-n}\ln\left[2\,\cosh\left(\frac{\alpha}{2\rm ch}-\frac{\nu}{2}\right)\right]
-(-1)^n(\nu\to-\nu),\nonumber\\
&&\nonumber\\
&&{\rm ch}=\cosh b, \qquad {\rm sh} = \sinh b, \qquad\hbox{we obtain}\nonumber\\
&&\nonumber\\
\left.I_0\right|_{\kk=1}\!\!&:&\left.-\partial_\nu h_0\right|_{\alpha=a\,\rm ch/sh},\nonumber\\
\left.I_1\right|_{\kk=1}\!\!&:&\left.-\partial_\nu h_2-\frac{1}{y}-\ln\alpha\right|_{\alpha=a\,\rm ch/sh},\nonumber\\
\lim_{\kk\to1}\ln\left(\frac{4}{\kk'}\right)F_2\!\!&:&\left.-\frac{h_1}{a}-\partial_\nu h_2
 +\frac{b\,\rm ch}{{\rm sh}^3}\,\partial_\nu h_0-1\right|_{\alpha=a\,\rm ch/sh},\nonumber\\
J_1\!\!&:&\left.{\rm sh\,ch}\,\partial_b\partial_\nu h_2\right|_{\alpha=a_0\,{\rm ch/sh},\,b=b_0}.
\end{eqnarray}

\end{document}